\documentclass{emulateapj}
\usepackage{txfonts}
\usepackage[utf8x]{inputenc}
\usepackage{graphicx}
\usepackage{natbib}
\usepackage{url}
\usepackage{amssymb}
\newcommand{\dMS}{$\Delta\log({\rm SFR})_{\rm MS}$}
\newcommand{\IRXbeta}{$A_{\rm IRX}$--$\beta$}

\slugcomment{Draft \today}
\shorttitle{The FIR--UV relation up to z=2.5}
\shortauthors{Nordon et al.}
\begin{document}
\title{The Far-Infrared, UV and Molecular Gas Relation in Galaxies up to z=2.5.}

\author{
R.~Nordon\altaffilmark{1},
D.~Lutz\altaffilmark{2},
A.~Saintonge\altaffilmark{2},
S.~Berta\altaffilmark{2},
S.~Wuyts\altaffilmark{2},
N.~M.~F\"orster~Schreiber\altaffilmark{2},
R.~Genzel\altaffilmark{2},
B.~Magnelli\altaffilmark{2},
A.~Poglitsch\altaffilmark{2},
P.~Popesso\altaffilmark{2},
D.~Rosario\altaffilmark{2},
E.~Sturm\altaffilmark{2},
L.~J.~Tacconi\altaffilmark{2}
}
\altaffiltext{1}{School of Physics and Astronomy, The Raymond and Beverly Sackler Faculty of Exact Sciences, Tel-Aviv University, Tel-Aviv 69978, Israel \email{nordon@astro.tau.ac.il}, \altaffilmark{2}{Max-Planck-Institut f\"ur Extraterrestrische Physik, Giessenbach 1, 85748 Garching, Germany}}

\begin{abstract}
We use the infrared excess (IRX) FIR/UV luminosity ratio to study the relation between the effective UV attenuation ($A_{\rm IRX}$) and the UV spectral slope ($\beta$) in a sample of 450 $1<z<2.5$ galaxies.
The FIR data is from very deep {\it Herschel} observations in the GOODS fields that allow us to detect galaxies with SFRs typical of galaxies with log($M_*$)$>$9.3.
Thus, we are able to study galaxies on and even below the main SFR--stellar~mass relation (main sequence).
We find that main sequence galaxies form a tight sequence in the IRX--$\beta$ plane, which has a flatter slope than commonly used relations.
This slope favors a SMC-like UV extinction curve, though the interpretation is model dependent.
The scatter in the \IRXbeta\ plane, correlates with the position of the galaxies in the SFR--$M_*$ plane.
Using a smaller sample of galaxies with CO gas masses, we study the relation between the UV attenuation and the molecular gas content.
We find a very tight relation between the scatter in the IRX--$\beta$ plane and the specific attenuation $S_A$, a quantity that represents the attenuation contributed by the molecular gas mass per young star.
$S_A$ is sensitive to both the geometrical arrangement of stars and dust, and to the compactness of the star forming regions.
We use this empirical relation to derive a method for estimating molecular gas masses using only widely available integrated rest-frame UV and FIR photometry.
The method produces gas masses with an accuracy between 0.12--0.16~dex in samples of normal galaxies between z$\sim$0 and z$\sim$1.5.
Major mergers and sub-millimeter galaxies follow a different $S_A$ relation.

\end{abstract}


\maketitle

%%%%%%%%%%%%%%%%%%%%%%%%%%%%%%%%%%%%%%%%%%%%%%%%
\section{Introduction} \label{sec:Introduction}
%%%%%%%%%%%%%%%%%%%%%%%%%%%%%%%%%%%%%%%%%%%%%%%%

The ultra-violet (UV) part in the spectrum of star forming galaxies (SFG), between the Lyman edge and $\sim$3000~\AA, is dominated by emission from young stars and provides strong constraints on the ongoing star formation rate (SFR).
The availability of many SFR tracers diminishes at moderately-high redshifts ($z>1$), where nebular lines are red shifted into the near-infrared and spectroscopy becomes increasingly difficult, while the availability of deep mid and far infrared (MIR and FIR) photometry is limited to a few deep extragalactic fields.
In contrast, at these redshifts the UV band shifts into the optical range, which makes it possible to observe from the ground to very high depths, often making it the most accessible SFR indicator for large samples.

UV radiation is susceptible to attenuation by dust in the host galaxy. In many examples, only a few percent of the UV photons escape the host galaxy.
It is therefore vital to correct UV luminosities for dust attenuation.
The estimation of the attenuation normally relies on measuring the change in some photometric colors, often at wavelengths other than the ones to be corrected. This estimation also requires assumptions on the extinction curve and the geometrical arrangement of the UV sources and dust.

In particular, the relation between the UV attenuation and the UV spectral slope $\beta$ ($f_\lambda \propto \lambda^\beta$) has been empirically calibrated \citep[e.g., ][]{Meurer99, Calzetti00}, mostly on local galaxies, and used in $z>1$ studies \citep[e.g., ][]{Daddi04}.
The calibrations used the ratio of the FIR luminosity (representing the UV radiation reprocessed by dust) to the UV luminosity (escaped light) as an indicator of the UV attenuation.
The FIR/UV ratio is often referred to as the infrared excess or IRX and we label the UV attenuation derived from the FIR/UV ratio as $A_{\rm IRX}$.

Since the works of \citet[][M99 hereafter]{Meurer99} and \citet[][C00 hereafter]{Calzetti00} on the \IRXbeta\ relation with the International Ultraviolet Explorer (IUE), Infrared Space Observatory (ISO) and Infrared Astronomical Satellite (IRAS) data, a large body of work has been written on the subject, for example:
\citet{Charlot00}, \citet{Kong04}, \citet{Buat05}, \citet{seibert05}, \citet{Howell10}, \citet{Hao11}, \citet{Wild11}, \citet{Overzier11}, \citet{Shimizu11}, and references therein.
The need for far infrared luminosities in order to determine $A_{\rm IRX}$ significantly limited earlier studies to low redshifts.
At redshifts of $z>1$ such studies were limited to very FIR-bright objects, or to use alternative methods for determining the UV attenuation.
The situation improved with the advent of {\it Spitzer} (used by some of the above references) and recently {\it Herschel} \citep{Pilbratt10} infrared space telescopes.
The latter in particular performed deep extragalactic surveys that detect normal star forming galaxies (SFGs) at redshifts of 2 and more with a wavelength coverage from 70 to 500 $\mu$m in six bands that cover the peak of the FIR emission in SFG.
Some of the recent {\it Herschel} results regarding the UV attenuation and $A_{\rm IRX}$--$\beta$ include: \citet{Nordon10}, \citet{Buat10}, \citet{wijesinghe11}, \citet{Burgarella11}, \citet{Buat11b}, \citet{Reddy12}.

Most of the above mentioned studies report significant scatter in the \IRXbeta\ relation, with different galaxy populations tending to lie on different relations.
This may indicate different dust properties, different stellar populations that dominate the UV, or different star and dust geometry in these populations.
The latter is of great importance as the arrangement of stars and dust is not yet clear and may have a strong effect on $A_{\rm IRX}$--$\beta$. Some authors favor a dust screen in the foreground with the UV emitting stars behind it.
Others find better agreement with a mix of dust and stars in the same volume, or combinations of both possibilities.
For example, \citet{Buat96} adopt a model where the stars and dust are mixed up to some scale height in the galaxy. Above this layer, additional dust acts as a foreground screen.
The model of \citet{Charlot00} incorporates an age-dependent obscuration where the younger stars are more obscured than older stars that already left, or cleared away the molecular clouds in which they formed.

A relatively tight relation has been found between the stellar mass of galaxies and their SFR.
This relation evolves with redshift and has been shown to be in place already at z=2 \citep{Brinchmann04, Noeske07, Elbaz07, Daddi07a}.
The relation is often referred to as the main-sequence of star forming galaxies.
In this paper we will refer to it as the `main sequence' (MS) for brevity.
Main sequence galaxies at a given redshift show many similarities to each other, in particular in properties that relate to star formation: e.g., they tend to show similar mid- to far-infrared SEDs \citep{Elbaz11, Nordon12}, SFR surface densities \citep{Elbaz11, Rujopakarn11, Wuyts11b}, gas depletion time scales \citep{Daddi10, Tacconi10, Saintonge11b}, light profile \citep{Wuyts11b}, etc.
These properties tend to change for galaxies with an enhanced SFR over the main relation (i.e. `above' the main sequence).
It has been known that galaxies with different `birthrate' parameters $b = SFR / <SFR>$ (ratio of current to past SFR, where $<SFR>$ is the mass over time since the initial formation of the galaxy)	 tend to lie on different $A_{\rm IRX}$--$\beta$ relations \citep[e.g., ][]{Kong04}.
At a given redshift, this parameter is closely related to the specific SFR (SSFR=$SFR/M_*$).
Since main sequence galaxies at a given redshift tend to have a narrow range of SSFR, one may expect an $A_{\rm IRX}$--$\beta$ relation for main sequence galaxies.
Such a relation is expected to have some scatter to it, in particular at high redshifts ($z \gtrsim 3$) where time scales are short and the star formation history can have a significant effect on the inferred physical quantities \citep{Schaerer12}.

Stars form inside molecular clouds and a relatively tight relation has been found between the molecular gas surface density and the SFR surface density \citep{Schmidt59, Kennicutt89}, often referred to as the Kennicutt-Schmidt (KS) relation.
Resolved giant molecular clouds (GMCs) in the Milky-Way and nearby galaxies show high optical depths and so the gas and dust in which the stars form also attenuate their UV emission.
One therefore may expect a relation between the SFR, the molecular gas and the attenuation.
However, there has been little evidence for a direct relation between the molecular gas and the UV attenuation, except for a general broad trend of increasing attenuation with the molecular gas surface density \citep{Buat96, Boissier07}.
\citet{Wuyts11b} used a simple model to predict the FIR to UV luminosity ratio from the SFR surface density. The model used the KS relation to derive the molecular gas surface density from the SFR. Then, with some assumptions on the geometry, they derived the column density for attenuation.
Their model was fairly successful at low redshifts, but developed increasing deviations as redshift increased.
The nature of such an attenuation--gas relation should hold some clues as to the geometrical arrangement of the star and attenuating material, yet it is still unclear.

In this study we use FIR photometry from the Photodetector Array Camera and Spectrometer \citep[PACS, ][]{Poglitsch10} on board {\it Herschel}, combined with UBVIz photometry
to study the $A_{\rm IRX}$--$\beta$ in $1<z<2.5$ galaxies.
When combined with the broad and deep optical coverage of these fields, the PACS observations at 70, 100 and 160 um, reaching depths and resolutions never available before, allow us to build a sizable sample to study the \IRXbeta\ relation and its scatter.
Thus, we are able to perform this study at the redshifts where rest-UV observations become highly useful and often the most accessible SFR indicator.
In addition, we study the relation between molecular gas content from CO measurements and the UV attenuation and spectral slope in $1<z<1.5$ normal star forming galaxies (SFG).
We also test these relations on low redshift galaxies.

The paper is structured as follows:
In section~\ref{sec:Data} we describe the data and samples used in this study.
In section~\ref{sec:A vs beta} we study the $A_{\rm IRX}$--$\beta$ diagram and the structure of the scatter in it.
In section~\ref{sec:specific attenuation} we find a relation between $A_{\rm IRX}$, $\beta$ and the molecular gas mass.
Then, in section~\ref{sec:M_gas_from_S_A} we derive a method to estimate molecular gas masses based on FIR and UV luminosities, which we test on a low redshift sample.
In section~\ref{sec:Discussion} we discuss the effect of various stars and dust configurations on the $A_{\rm IRX}$--$\beta$ relation and the reason we can determine gas mass from FIR and UV photometry.
Our conclusions are summarized in section~\ref{sec:conclusions}.

Throughout this paper we assume a \citet{Chabrier03} IMF, and a cosmology with ($\Omega_m$,$\Omega_\Lambda$,$H_0) = (0.3,0.7,70$~km~s$^{-1}$~Mpc$^{-1}$).

%%%%%%%%%%%%%%%%%%%%%%%%%%%%%%%%%%%%%%
\section{Data and Samples} \label{sec:Data}
%%%%%%%%%%%%%%%%%%%%%%%%%%%%%%%%%%%%%%

\begin{table}
\begin{center}
\caption{\label{tab:PEP fields} PACS photometry depths in the fields used in this study.}
\begin{tabular}{@{}lccc}
\hline
      & \multicolumn{3}{c}{3$\sigma$ flux limits [mJy]}\\
Field & 70 $\mu$m & 100 $\mu$m & 160 $\mu$m\\
\hline
GOODS-N & \dots & 0.93 & 2.1 \\
GOODS-S & 0.81   & 0.51 & 1.26 \\
\hline
\end{tabular}
\end{center}
\end{table}

Our far-infrared data is based on {\it Herschel}-PACS observations in the 
GOODS fields.
We combine the data from two {\it Herschel} large programs: the PACS Evolutionary Probe (PEP\footnote{\url{http://www.mpe.mpg.de/ir/Research/PEP/}}) project, a guaranteed time program \citep{Lutz11} and the GOODS-Herschel open time key program \citep{Elbaz11}.
The combined reduction is described in details in \citet{Magnelli12} and is similar in principle to the reduction of the PEP data as described in \citep{Lutz11}.
The limiting fluxes are listed in Table~\ref{tab:PEP fields}.
The PACS 160, 100 and 70 $\mu$m fluxes were extracted using sources 
from a {\it Spitzer} MIPS 24~$\mu$m catalog as priors, following the method 
described in \citet{Magnelli09} (see also \citet{Lutz11} for more details).
For our sample selection we require a redshift (either photometric or 
spectroscopic), a stellar mass, a PACS detection, and sufficient optical bands to derive rest frame 1600~\AA\ and 2800~\AA\ magnitudes.
Below we describe the constraints each requirement imposes on our sample selection.

Photometric redshifts are the same as used in \citet{Nordon12} and thanks to the large number of bands and the high limiting magnitudes, their quality is very good. When compared to the spectroscopic redshifts: $(z_{phot}-z_{spec})/(1+z_{spec})$ = 0.033 and 0.038 in GOODS-N and GOODS-S respectively.
63\% and 52\% of the selected galaxies in GOODS-S and GOODS-N respectively have spectroscopic redshifts.

The stellar masses are the same as used in \citet{Nordon12}.
They were calculated using the method described in 
\citet{Fontana04}, with adjustments as described in \citet{Santini09},
see also \citet{Fontana06} and \citet{Grazian06}.
To ensure good mass estimates, we require a detection in all of the 3.6--5.8 $\mu$m IRAC bands 
that cover the SED rest frame stellar bump (rest frame H-band) in $z>1$ galaxies.
We place a conservative lower limit on the selected galaxies masses of $\log(M_*/M_\odot)>9.3$ in order to avoid the galaxies with the lowest derived masses that are more likely to be critical errors in the mass fit.
This limit ensures detections in most of the bands between U and IRAC 5.8~$\mu$m and thus reliable mass estimates.

$L_{\rm IR}$(8--1000~$\mu$m) is calculated by fitting \citet[][CE01 hereafter]{CE01} templates to the available PACS fluxes. Both template shape and scale are allowed to vary and the solution with minimal $\chi^2$ value is selected.
To derive errors on $L_{\rm IR}$ we randomize the observed fluxes according to their errors while assuming a normal distribution and repeat the template fitting to derive a new luminosity.
This step is repeated a large number of times and the standard deviation (STDEV) of the measured luminosities is adopted as the error on $L_{\rm IR}$.
If only one PACS band is available (16\% of the galaxies, virtually all with only 160~~$\mu$m), we fit a CE01 template while preserving the original scaling of the templates.
This introduces only a small additional error and no significant bias at redshifts $1.0<z<2.5$ (see \citet{Elbaz10} and the appendix in \citet{Nordon12}).

We estimate rest frame 1600~\AA\ and 2800~\AA\ magnitudes ($M_{1600}$ and $M_{2800}$ respectively) by interpolating between the available bands.
We use UBVIz photometry from various instruments.
In GOODS-N a photometric catalog has been extracted and compiled by S.~Berta for the PEP collaboration \citep[used in papers such as ][and others]{Wuyts11b, Nordon12} and has been publicly released
\footnote{\url{http://www.mpe.mpg.de/ir/Research/PEP/public\_data\_releases.php}}.
The GOODS-S optical photometry is from the MUSIC catalog \citep{Santini09}.
We exclude bands that have their central wavelengths between 2000--2300~\AA\ in order to avoid the bands from being dominated by an enhanced attenuation feature around 2170~\AA\ (the so called `UV bump') that may \citep{Buat11b} or may not \citep{Calzetti94} exist at $z>1$.
To derive $M_{1600}$ we use all remaining filters between rest 1200--2800~\AA\ and interpolate the flux at 1600~\AA\ (converted to AB magnitudes) by fitting a power-law.
The fitting minimizes a weighted $\chi^2$, where we weight the magnitude errors by $1+|\lambda_{\rm filter}-1600| / 1600$~\AA. This is done in order to give more weight to filters that observe close to the desired wavelength.
Errors are estimated by randomizing the input fluxes according to their respective errors, while assuming a normal distribution and interpolating again to obtain $M_{1600}$. The process is repeated a large number of times and the standard deviation on the derived $M_{1600}$ is adopted as the error.
To derive $M_{2800}$ we select the filters that observe rest 1500--3500~\AA\ (again excluding the range of the `UV bump') and interpolate in the same way as we did with $M_{1600}$.
To ensure a reliable interpolation to the two wavelengths we require at least one band that observes less than a full filter width from the desired rest frame wavelength and at least two bands in total.
For this reason we limit our lowest redshift to $z>1$, where 1600~\AA\ rest is covered by the U-band and above.
Finally, we select only galaxies for which the final error on $\beta$ is lower than 0.2.

Since the presence of an active galactic nuclear (AGN) can significantly affect the emission in UV, we remove galaxies suspected as AGN hosts from the sample.
Sources which have an X-ray counterpart in the Chandra 2~Ms catalogs 
\citep{Alexander03, Luo08} have conservatively been flagged as suspected AGN hosts and removed.
None of the removed galaxies clearly comply with the \citet{Bauer04} classification as star forming galaxies according to their X-ray emission.
While we may be removing some exceptionally rapid star forming galaxies, whose X-ray emission is due to the star formation and not the AGN \citep[e.g.,][]{Alexander05}, it will have a negligible effect on this study that focuses more on the bulk of the galaxy population around the main-sequence and a little above it.
AGNs that are obscured enough to prevent detectable X-ray emission will also have their UV emission highly attenuated and the observed UV can be attributed to the star formation.

\begin{figure}[t]
 \centering
 \includegraphics[width=\columnwidth, clip=true, trim=80pt 0pt 80pt 0pt]{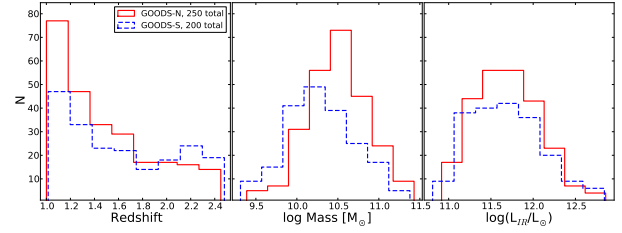}
 \caption{Several properties of the sample:
  {\it left}: The redshift distribution. {\it Center}: Stellar mass distribution. {\it Right}: Total infrared luminosity distribution.}
\label{fg:sample_histograms}
\end{figure}

In total, our sample includes 250 and 200 sources in GOODS-N and GOODS-S respectively.
The properties of our sample in redshift, mass and $L_{\rm IR}$ distributions are presented in figure~\ref{fg:sample_histograms}.

%%%%%%%%%%%%%%%%%%%%%%%%%%%%%%%%%%%%%%%%%%%%%%%%%%%%%%%%%%%%%%
%
\section{The Effective UV Attenuation Versus UV Spectral Slope}
\label{sec:A vs beta}
%
%%%%%%%%%%%%%%%%%%%%%%%%%%%%%%%%%%%%%%%%%%%%%%%%%%%%%%%%%%%%%%

\subsection{Terms and Definitions}
\label{sec:terms_and_definitions}

Star formation rates can be calculated both from the total 8--1000 $\mu$m infrared luminosity and from the UV luminosity at 1600~\AA.
We adopt the conversions of \citet{Kennicutt98}, modified downward by a factor of 1.6 to match a \citet{Chabrier03} initial mass function (IMF):
\begin{equation}
 \frac{SFR_{\rm IR}}{\rm M_\odot yr^{-1}} = 1.09 \times 10^{-10} \frac{L_{\rm IR,\,(8-1000\,\mu{\rm m})}}{L_\odot}
\end{equation}
\begin{equation}
 \frac{SFR_{\rm UV}}{\rm M_\odot yr^{-1}} = 0.875\times 10^{-28} \frac{L_{\nu,\,1600\,\AA}}{{\rm erg\,s^{-1}\,Hz^{-1}}}
\end{equation}
The conversion from $L_{\rm IR}$ to SFR assumes that $L_{\rm IR}$ represents the bolometric luminosity of the young stars. In practice, we detect UV emission, indicating that some radiation escapes and is not reprocessed into the infrared.
Therefore, for the total SFR we use: $SFR_{\rm total} = SFR_{\rm IR} + SFR_{\rm UV}$.

$SFR_{UV}$ uses the observed UV luminosity and hence is missing a significant fraction of the luminosity which was attenuated by the dust.
The effective UV attenuation $A_{\rm IRX}$ should correct $SFR_{UV}$ to $SFR_{total}$, i.e. $\log(SFR_{\rm total}) = \log(SFR_{\rm UV}) + 0.4A_{\rm IRX}$. 
We define $A_{\rm IRX}$ as:
\begin{equation}
 A_{\rm IRX} = 2.5 \log \left( \frac{SFR_{\rm IR}}{SFR_{\rm UV}} + 1 \right)
 \label{eq:A_IRX_definition}
\end{equation}

Let $\beta$ be the spectral slope, such that $f_\lambda \propto \lambda^\beta$. While spectra between two wavelengths do not always resemble a perfect power law, $\beta$ can always be formally defined between two wavelengths as:
\begin{equation}
 \beta = \frac{ \log(f_{\lambda_1}/f_{\lambda_2}) }{ \log(\lambda_1/\lambda_2) } = \frac{ -0.4(M_1-M_2) }{ \log(\lambda_1/\lambda_2) } -2
\label{eq:beta_definition}
\end{equation}
where $M_1$ and $M_2$ are the AB magnitudes at wavelengths $\lambda_1$ and $\lambda_2$ respectively.
In this study we adopt $\lambda_1 = 1600$~\AA, $\lambda_2 = 2800$~\AA\ and obtain rest frame $M_1$ and $M_2$ from the available photometry as explained in Section~\ref{sec:Data}.

%%%%%%%%%%%%%%%%%%%%%%%%%%%%%
\subsection{$A_{\rm IRX}$--$\beta$ for The PACS Selected Sample}
%%%%%%%%%%%%%%%%%%%%%%%%%%%%%

\begin{figure}[t]
 \centering
 \includegraphics[width=\columnwidth, clip=true, trim=15pt 0pt 30pt 0pt]{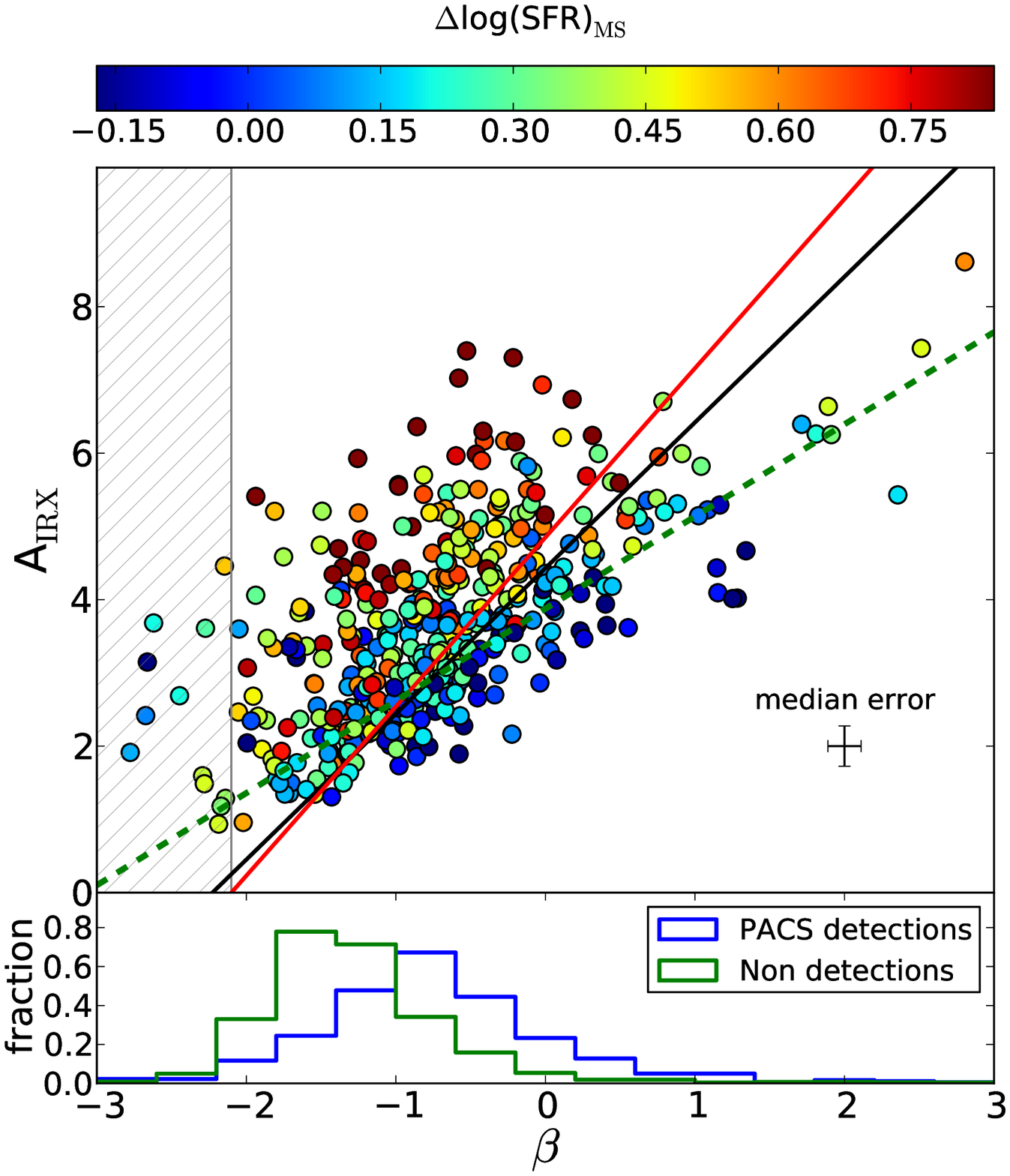}
 % Abeta_dMScolors.eps: 0x0 pixel, 300dpi, 0.00x0.00 cm, bb=13 175 598 616
 \caption{The effective UV attenuation versus UV spectral slope for $1.0<z<2.5$ PACS detected galaxies.
The color scale indicates the $\Delta\log({\rm SFR})_{\rm MS}$ distance of the galaxy from the main-sequence.
Lines indicate various $A_{\rm IRX}$--$\beta$ relations: solid black - \citet{Meurer99}, solid red - \citet{Calzetti00}, dashed green - best fit for galaxies with $-0.15 < \Delta\log({\rm SFR})_{\rm MS} < 0.15$.
The hatched area indicates expected values of $\beta$ for an unobscured population.
The lower panel shows the distribution in $\beta$ of the PACS detected sample (blue). The distribution of sources with similar selection criteria, but without a PACS detection is plotted in green.}
 \label{fg:A-beta dMS colors}
\end{figure}

In Figure~\ref{fg:A-beta dMS colors} we plot the $A_{\rm IRX}$--$\beta$ relation for the full $1<z<2.5$ PACS-detected sample.
There is a significant scatter in the relation. However, the galaxies were color-coded according to their distance from the main-sequence in \dMS\ and the color scale reveals a structure to this scatter.
We define \dMS~$=\log( {\rm SFR}_{\rm galaxy} / {\rm SFR}_{\rm MS} )$, where ${\rm SFR_{MS}(M_*,z)}$ is the SFR on the main sequence for the given mass and redshift. We use the \citet{Rodighiero10} redshift-dependent main-sequence curves. These were derived from {\it Herschel}-PACS data in GOODS-N, partial to the data set used here.
For convenience, we give the main sequence curves below:
\begin{equation}
\log\left( \frac{SSFR}{\rm Gyr^{-1}} \right) = 
\left\{
\begin{array}{lr}
 -0.28 \log\left( \frac{\rm M_*}{\rm M_\odot} \right) +2.57; & z=0.75 \\
 -0.51 \log\left( \frac{\rm M_*}{\rm M_\odot} \right) +5.32; & z=1.25 \\
 -0.50 \log\left( \frac{\rm M_*}{\rm M_\odot} \right) +5.25; & z=1.75\\
\end{array}
\right.
\label{eq:the_main_sequence}
\end{equation}
%We adopt the central redshift of each redshift bin in \citet{Rodighiero10}.
To find the main sequence at a specific redshift we linearly interpolate between the above curves for the given mass. For $1.75<z<2.5$ we assume that the main sequence does not evolve beyond $z=1.75$.

\begin{figure}[t]
 \centering
 \includegraphics[width=\columnwidth]{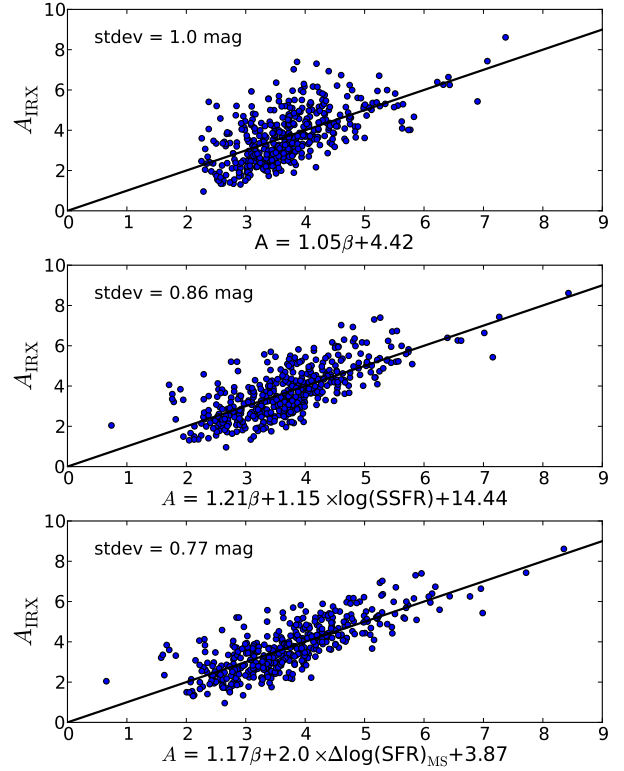}
 % A_correlations.eps: 0x0 pixel, 300dpi, 0.00x0.00 cm, bb=13 30 598 761
 \caption{The UV attenuation from IRX relation versus the best combination of $\beta$ and SSFR or \dMS.
{\it Top:} Best fit for the entire sample with no additional parameters.
{\it Center:} When including SSFR as a parameter in the fit.
{\it Bottom:} When including \dMS\ as a parameter in the fit.
The standard deviation in attenuation magnitudes from the 1:1 relation is indicated in each panel.}
 \label{fg:planes fit}
\end{figure}

It is evident that the higher the \dMS\ of the galaxy, the higher it tends to lie on the $A_{\rm IRX}$--$\beta$ relation.
As mentioned in the introduction, local galaxies show a very similar effect in the `birth-rate' parameter, closely related to the SSFR. The main difference between SSFR dependency and \dMS\ dependency is that constant SSFR is independent of the galaxy mass, while constant \dMS\ is mass dependent due to the non-zero slope of the main sequence (e.q.~\ref{eq:the_main_sequence}).
To verify that indeed the slope of the main-sequence has an effect and that \dMS\ is the more important parameter for the scatter, we fit a plane to the data in $A_{\rm IRX}$, $\beta$ and either SSFR or \dMS\ space.
The result is plotted in Figure~\ref{fg:planes fit}.
As a first order correction, \dMS\ allows a better minimization of the scatter than simply using the SSFR.
Most of the effect is due to the slope and not so much due to the absolute scale of the main-sequence, since the adopted main-sequence of \citet{Rodighiero10} evolves little from $z=1$ to $z=2$, but has a significant slope of $\sim$-0.5 in SSFR versus mass.
We warn that the relations in Figure~\ref{fg:planes fit} should not be used to derive $A_{\rm IRX}$. Such attempts will result in a significant fraction of large attenuation correction errors.
We also expect systematic deviations, in particular  for galaxies with $\beta \lesssim -1.5$ due to the dependency on \dMS\ that may be non-linear over the full range of $\beta$.

Galaxies on the main-sequence form a near straight sequence in $A_{\rm IRX}$--$\beta$ (Figure~\ref{fg:A-beta dMS colors}).
We fit a linear relation to the galaxies close to the main-sequence ($-0.15<$\dMS$<0.15$) while minimizing the $\chi^2$ for the orthogonal projection of the $A_{\rm IRX}$ and $\beta$ errors and derive:
\begin{equation}
 A_{\rm MS} = 1.26\beta + 3.90
\label{eq:A_MS}
\end{equation}
This slope is shallower than the relations derived by M99 and C00 (Figure~\ref{fg:A-beta dMS colors}).
Taken at face value and assuming that the obscuring material is arranged as a screen in front of the sources, an $A_{\rm IRX}$--$\beta$ slope of $\alpha_{\rm IRX}=1.26$ translates into a dust-extinction law with $\lambda^{-1.18}$ wavelength dependency (e.q.~\ref{eq:A_IRX_slab} in the Appendix).
Such a UV extinction law is almost identical to the Small Magellanic Cloud (SMC) extinction law \citep{Gordon03}, that has $A(\lambda) \propto \lambda^{-1.15}$ between 1600 and 2800~\AA.

It is clear that the way the sample is selected can significantly affect the derived $A_{\rm IRX}$--$\beta$ relation. For example, a high FIR luminosity cut will tend to select galaxies above the main-sequence (in SSFR), resulting in a higher and possibly steeper $A_{\rm IRX}$--$\beta$ relation.
Mixing different redshifts can result in different \IRXbeta\ relation since galaxies with ULIRG-like IR luminosities have high \dMS\ at z$\sim$0, but can be main sequence galaxies at $z>1$. 
\IRXbeta\ samples that are selected to span a large range in $\beta$ will tend to select fundamentally different galaxies at the two extremes of the $\beta$ range. 

We also tested a subsample of BzK selected \citep{Daddi04} galaxies from our full sample. The picture for the BzKs does not change significantly from Figure~~\ref{fg:A-beta dMS colors}. This is somewhat expected because almost all $z>1.4$ galaxies in our sample are also selected as BzKs, when the Ks imaging is deep enough.
The M99, C00 and $A_{MS}(\beta)$ relations cross each other around $\beta \approx -1$, which is also the peak of the number distribution in $\beta$ (Figure~~\ref{fg:A-beta dMS colors} lower panel). The peak in the $\beta$ number distribution for sources undetected by PACS is at a slightly lower $\beta$.
This means that for the large majority of UV galaxies, all 3 relations will produce similar attenuation corrections. 
However, biases will develop depending on the assumed \IRXbeta\ relation and the sample selection, in particular for galaxies with much higher/lower $\beta$ than -1.

In the next sections, we will study in more details the scatter in the \IRXbeta\ relation, the parameters governing it, and the information it provides about the internal structure of the galaxies.

%%%%%%%%%%%%%%%%%%%%%%%%%%%%%%%%%%%%%%
%
\section{Molecular Gas and The Specific Attenuation}
\label{sec:specific attenuation}
%
%%%%%%%%%%%%%%%%%%%%%%%%%%%%%%%%%%%%%%

\begin{figure}[t]
 \centering
 \includegraphics[width=\columnwidth]{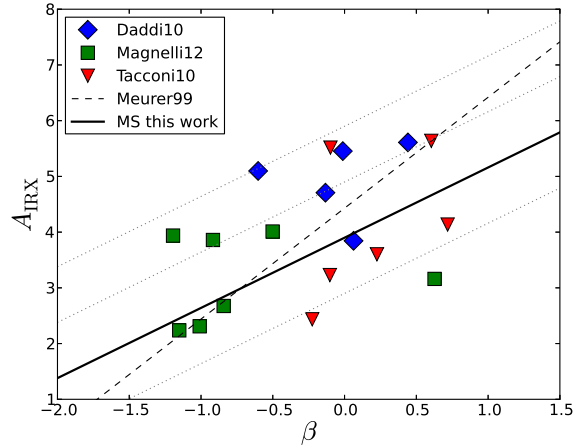}
 % CO_sample_IRXbeta.eps: 0s pixel, depict,  cm, BB=13 175 598 616
 \caption{The $A_{\rm IRX}$--$\beta$ relation for the CO sample. Colors and symbols indicate the literature source: blue diamonds are from \citet{Daddi10}, red triangles are from \citet{Tacconi10}, green squares are from \citet{Magnelli12}. The dotted line mark the \citet{Meurer99} relation. The solid black line indicate the relation for the main-sequence galaxies from this work and dotted lines are spaced 1 magnitude difference from the main-sequence relation.}
 \label{fg:A-beta for CO}
\end{figure}

Stars form in molecular clouds and a tight relation is established between the SFR and the molecular mass surface density \citep{Schmidt59, Kennicutt89}, often referred to as the `Kennicutt--Schmidt' (KS) relation.
Molecular clouds contain dust that is mixed with the gas and makes them optically thick to UV, obscuring the stars forming in it.
The relation between the optical depth and the molecular gas content (with the dust embedded in it) strongly depends on the geometrical arrangement of the stars and dust.
Geometries will be discussed in detail in Section~\ref{sec:Discussion}, but for the moment we can consider the following simple example: (a) two stars and two dust elements, each star with a dust element in front of it that produce a certain attenuation, versus (b) two stars crowded behind a single dust element that obscures both to the same attenuation value as the former case.
Therefore, (a) requires more gas and dust per obscured star than (b) to produce the same attenuation.

Let us define a new quantity, the `specific attenuation' ($S_A$) as:
\begin{equation}
 S_A \equiv \tau_{UV} \frac{ A_{\rm IRX}{\rm SFR} }{ M_{mol} } \propto \frac{A_{IRX}}{M_{mol}/N_{UV}}
 \label{eq:S_A_defintion}
\end{equation}
%%%
$S_A$ is proportional to the attenuation per unit molecular gas available per UV emitting star.
However, the total number of UV sources $N_{UV}$ is not a convenient quantity to use.
The calibration of the UV luminosity as a SFR indicator relies on the assumption that the UV emitting population reaches a steady state in less than $10^8$~years of continuous star formation \citep{Kennicutt98}.
Hence, the SFR in the above equation serves as a more practical proxy to the number of UV emitting stars ($N_{UV}$).
$\tau_{UV}$ is a constant with time dimensions that makes $S_A$ dimensionless.
The exact value of $\tau_{UV}$ is of little consequence in this study as it will be blended by subsequent empirical fitting. 
Purely for scaling convenience, we will use $\tau_{UV}=10^9$~yr.
Thus, for the typical SFR/$M_{mol} \sim 10^{-9}$~yr$^{-1}$ and $A_{\rm IRX} \sim 1$, we expect $S_A$ in order of unity.
$S_A$ tells us how efficiently the gas (and the dust embedded in it) is arranged in order to block the light from the stars.

$S_A$ strongly depends on the typical stars--dust geometrical arrangement (see Section~\ref{sec:Discussion}) and also varies with compactness.
If we keep the number of young stars and amount of dust constant, and increase the {\it local} surface densities by compressing the dimensions of the star forming regions, then the column density will increase, resulting in a higher $A_{IRX}$ and a higher $S_A$.
The physical meaning is that the stars are more crowded behind every gas (dust) element and each gas element blocks the lines of sight to more UV sources.
Whether the stars are behind the gas or mixed with it will also have a strong effect on $S_A$.

To derive $S_A$ we require CO observation of galaxies at $z>1$, that are also detected in the FIR.
For the moment, we would like to avoid the extreme and rare cases of sub-millimeter galaxies (SMGs) and concentrate on more common galaxies near the main-sequence.
A handful of such PACS-detected galaxies have CO measurements from the Plateau de Bure Interferometer (PdBI) in GOODS-N and the Extended Groth Strip (EGS) fields.
Table~\ref{tab:CO sample} specifies the details of our CO sample.
The sources in GOODS-N were already included in our main \IRXbeta\ sample and here we add the \citet{Tacconi10} sources in the EGS field, also observed with {\it Herschel}-PACS as part of the PEP project \citep{Lutz11}.
We correct the literature $M_{mol}$ to use a uniform conversion factor $\alpha_{\rm CO}=4.35$~$M_\odot$~K$^{-1}$~km$^{-1}$~s~pc$^{-2}$, such that $M_{mol} = \alpha_{\rm CO} \cdot L_{CO\, 1-0}$, where $L_{CO\, 1-0}$ is in K~km~s$^{-1}$~pc$^{2}$ and $M_{mol}$ is in $M_\odot$. $\alpha_{\rm CO}$ already includes a factor of 1.36 that accounts for the He mass.
The optical photometry that is used in deriving $\beta$ and 1600~\AA\ flux (same method as described in Section~\ref{sec:Data}) is from the public catalogs of the AEGIS\footnote{\url{http://aegis.ucolick.org/astronomers.html}} team.

In Figure~\ref{fg:A-beta for CO} we plot the \IRXbeta\ diagram for the CO sample. 
As can be seen, this sample is widely scattered around the \citet{Meurer99} relation and the $A_{MS}(\beta)$ (e.q.~\ref{eq:A_MS}) relation discussed above.
The sample has $\beta > -1.5$ for all the galaxies and avoids the cases of blue obscured galaxies, that suggest a bimodal distribution in $f_{\lambda_1}(A)$ (Section~\ref{sec:Discussion}).

\begin{figure}[t]
 \centering
 \includegraphics[width=\columnwidth]{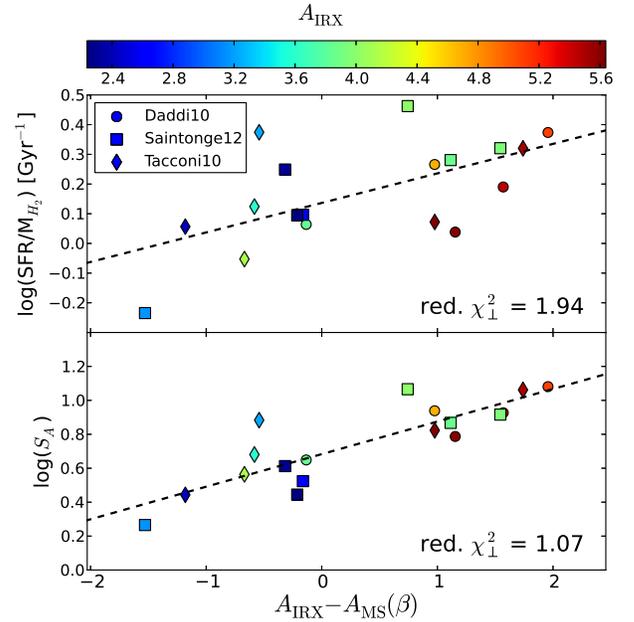}
 % specific_attenuation.eps: 0s pixel, depict,  cm, BB=13 30 598 761
 \caption{{\it Top:} The SFE ($\tau_{dep}^{-1}$) versus the attenuation distance from the $A_{\rm IRX}$--$\beta$ main-sequence relation.
 Colors indicate the $A_{\rm IRX}$ for the galaxies matching the color bar at the top.
 Symbols indicate the source for the gas masses.
 The reduced perpendicular-$\chi^2$ is indicated in the lower right corner.
 {\it Bottom}: The same as above for the specific attenuation $S_A$.}
 \label{fg:specific attenuation}
\end{figure}

In section~\ref{sec:A vs beta} and figure~\ref{fg:A-beta dMS colors} we established that the location on the $A_{\rm IRX}$--$\beta$ diagram is shifted upwards in $A_{\rm IRX}$ with increasing \dMS. \citet{Saintonge11b} found a relatively tight correlation between the gas depletion time-scale $\tau_{dep}=M_{mol}/{\rm SFR}$ and SSFR for local galaxies.
This relation scales with redshift like the main-sequence, which means that more generally, a decreasing $\tau_{dep}$ correlates with an increasing \dMS\ \citep{Saintonge12}.
In figure~\ref{fg:specific attenuation} top panel we plot SFR/M$_{mol}$ ($\tau_{dep}^{-1}$) as a function of the difference between the measured $A_{\rm IRX}$ (e.q.~\ref{eq:A_IRX_definition}) and $A_{\rm MS}(\beta)$ (e.q.~\ref{eq:A_MS}).
A clear correlation exists, which reflects the fact that \dMS\ and $\tau_{dep}$ are well correlated.

In the bottom panel of Figure~\ref{fg:specific attenuation} we plot $S_A$ as a function of $A_{\rm IRX} - A_{\rm MS}(\beta)$.
A tightening of the relation relative to the top panel is evident.
Note that although we are using $A_{\rm MS}$, only the slope in $A_{\rm MS}$ is important since the zero order terms are adjusted when fitting a linear relation to Figure~\ref{fg:specific attenuation}.
The best-fit linear relation is:
\begin{equation}
 \log(S_A) = 0.20_{\pm 0.04} [A_{\rm IRX} - 1.26(\beta-\beta_0)] + 0.46_{\pm 0.03}
 \label{eq:bestfit_SA}
\end{equation}
where $\beta_0 = -2.2$. The term $A_{\rm screen} \equiv 1.26(\beta-\beta_0)$ in the above equation presents $S_A$ as a function of the $A_{\rm IRX}$ excess over the attenuation of a uniform screen model for the given $\beta$.
The reasons for presenting $S_A$ in such a way are discussed in Section~\ref{sec:S_A_geometry_discussion}.

While the correlation appears to be tight, the values in the x and y axes are correlated.
In the bottom panel in particular, $A_{\rm IRX}$ explicitly appears on both axes.
To verify that the tighter relation in the bottom panel versus the top panel is indeed significant, we perform a $\chi^2$ test that takes correlations into account (denoted $\chi_\perp^2$ hereafter).
A full description of the test is given in Appendix~\ref{app:chisqr_test}.
The $\chi_\perp^2$ values are indicated in each panel and indeed $S_A$ offers a significant improvement in tightening the relation.
The $\chi_\perp^2$ for the $S_A$ relation is consistent with the expected from pure random noise of the flux measurements.
Such a tight relation is non trivial.
There are many possible arrangements of the stars and dust that will produce very different $S_A$, while still having the same location in the \IRXbeta\ plane, yet this is not the case in our sample.

The increase in $S_A$ with increasing $A_{\rm IRX}$ at a given $\beta$ indicates that the shift of the galaxies on the $A_{\rm IRX}$--$\beta$ plane is mostly due to geometry and not the unattenuated spectral slope $\beta_0$.
There is no compelling reason for $\beta_0$ to be tightly correlated with $S_A$.
Variation in the extinction law due to different dust grains between galaxies can in principle affect both $S_A$ (through $A/M_{mol}$) and $A_{\rm IRX}$ at a given $\beta$, to produce a correlation.
However, $S_A$ in our sample changes by factor $\sim$6,  which is more than what reasonable changes to the UV extinction law can produce.
We conclude that the scatter in the $A_{\rm IRX}$--$\beta$ relation is dominated by the geometrical arrangement of dust and stars as reflected by the values of the specific attenuation $S_A$.
Other factors considered such as $\beta_0$ and the UV extinction law seem to play a less significant role.

%% CO SAMPLE TABLE %%
\begin{table*}[t]
 \begin{center}
  \caption{\label{tab:CO sample} The $z\sim1$ sample of normal star forming galaxies with CO measurements and Herschel-PACS fluxes.}
    \begin{tabular}{l*{7}{c}}
     ID & Source$^\dagger$ & z & $\beta$ & $A_{\rm IRX}$ & $L_{\rm IR}^\ast$    & SFR$_{\rm tot}^\ddagger$                & $M_{mol}^\star$\\
        &        &   &         & mag           & $10^{12} L_\odot$ & $M_\odot$yr$^{-1}$ & $10^{10} M_\odot$\\
     \hline
     \hline
	BzK-4171 & D10 & 1.465& 0.44 $\pm$0.11& 5.6$\pm$0.2& 0.92$\pm$0.06& 101$\pm$7 & 9.3$\pm$2.2\\
	BzK-21000& D10 & 1.523& -0.60$\pm$0.10& 5.1$\pm$0.2& 2.10$\pm$0.08& 231$\pm$9 & 9.8$\pm$2.4\\
	BzK-16000& D10 & 1.522& 0.06$\pm$0.06 & 3.8$\pm$0.2& 0.73$\pm$0.11&  82$\pm$12& 7.1$\pm$1.7\\
	BzK-17999& D10 & 1.414& 0.76 $\pm$0.12& 6.5$\pm$0.2& 4.11$\pm$0.38 & 450$\pm$41& 7.6$\pm$1.8\\
	BzK-12591& D10 & 1.6  & -0.13$\pm$0.05& 4.7$\pm$0.2& 2.41$\pm$0.10& 267$\pm$11 & 14.5$\pm$3.5\\
	EGS 13004291& T10 & 1.2 & -0.10$\pm$0.06& 5.5$\pm$0.2& 5.32$\pm$0.14& 585$\pm$15& 28.0$\pm$1.1\\
	EGS 12007881& T10 & 1.17& -0.22$\pm$0.05& 2.4$\pm$0.2& 0.77$\pm$0.15& 94$\pm$16 & 8.3$\pm$0.4\\
	EGS 13017614& T10 & 1.18& 0.72$\pm$0.13 & 4.1$\pm$0.3& 0.74$\pm$0.15& 82$\pm$16 & 9.3$\pm$0.7\\
	EGS 13004661& T10 & 1.19& -0.10$\pm$0.1 & 3.2$\pm$0.4& 0.49$\pm$0.16& 57$\pm$17 & 2.4$\pm$0.5\\
	EGS 13003805& T10 & 1.23& 0.60$\pm$0.15 & 5.6$\pm$0.2& 1.93$\pm$0.21& 212$\pm$23& 18.0$\pm$1.2\\
	EGS 13011439& T10 & 1.1 & 0.23$\pm$0.09 & 3.6$\pm$0.3& 0.53$\pm$0.12& 60$\pm$13 & 4.5$\pm$1.0\\
	GN21172& M12 & 1.249& -0.50$\pm$0.12& 4.0$\pm$0.1& 0.42$\pm$0.04& 47$\pm$4 & 1.62$\pm$0.44\\
	GN17022& M12 & 1.224& -0.84$\pm$0.09& 2.7$\pm$0.1& 0.31$\pm$0.04& 37$\pm$4 & 2.95$\pm$0.68\\
	GN1464 & M12 & 1.084& -1.15$\pm$0.17& 2.2$\pm$0.1& 0.36$\pm$0.05& 45$\pm$5 & 3.63$\pm$0.91\\
	GN5532 & M12 & 1.021& -1.19$\pm$0.17& 3.9$\pm$0.1& 0.54$\pm$0.03& 60$\pm$3 & 2.88$\pm$0.95\\
	GN26843& M12 & 1.016& -0.92$\pm$0.17& 3.9$\pm$0.1& 0.98$\pm$0.04& 110$\pm$3& 5.75$\pm$1.38\\
	GN13168& M12 & 1.016& -1.01$\pm$0.17& 2.3$\pm$0.1& 0.41$\pm$0.03& 51$\pm$3 & 2.88$\pm$0.68\\
	GN177  & M12 & 1.17 & 0.63$\pm$0.11 & 3.2$\pm$0.2& 0.26$\pm$0.04& 30$\pm$4 & 5.13$\pm$0.72\\
    \hline
    \multicolumn{8}{l}{$^\dagger$ D10 - \citet{Daddi10}, T10 - \citet{Tacconi10}, M12 - \citet{Magnelli12}  } \\
    \multicolumn{8}{p{11cm}}{$^\ast$ from Herschel-PACS data of the PEP project (T10 sources) and the combined PEP \& GOODS-Herschel data (D10 and S12 sources).}\\
    \multicolumn{8}{l}{$^\ddagger$ SFR$_{\rm tot}$ = SFR(FIR) + SFR(UV)$_{\rm uncorrected}$ }\\
%     \multicolumn{8}{p{11cm}}{$^\star$ The mass of H$_2$+He from the literature source. Corrected for a uniform a conversion factor from $L_{\rm CO\, 1-0}$~[K~Km~s$^{-1}$] to $M_{mol}$ of 4.35. This already includes a factor of 1.36 that accounts for He mass in the molecular phase. }\\
    %\vspace{0.5cm}
    \end{tabular}
 \end{center}
\end{table*}

%%%%%%%%%%%%%%%%%%%%%%%%%%%%%%%%%%%%%%%%%%%%%%%%%%%%%%%%%%%%%
%
\section{Estimating Gas Masses With UV and FIR Photometry}
\label{sec:M_gas_from_S_A}
%
%%%%%%%%%%%%%%%%%%%%%%%%%%%%%%%%%%%%%%%%%%%%%%%%%%%%%%%%%%%%%

\subsection{The $S_A$ Method}

\begin{figure}[t]
 \centering
 \includegraphics[width=\columnwidth]{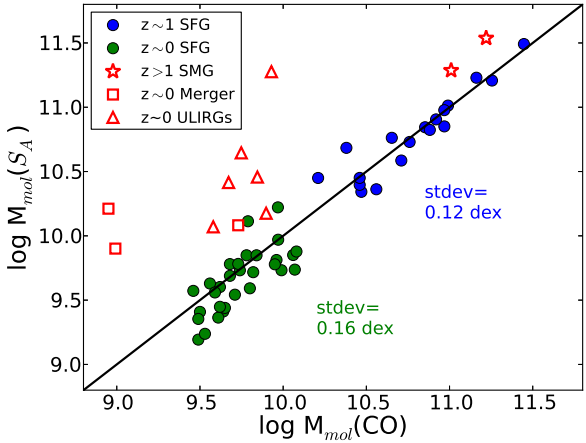}
 % Mgas_SA_vs_CO.eps: 0s pixel, depict,  cm, BB=13 175 598 616
 \caption{The $M_{mol}(S_A)$ as derived from the specific attenuation ($S_A$) relation, versus the $M_{mol}(CO)$ measured from CO observations.
 The solid line is the one to one relation.
 Blue circles are for the $z\sim1$ SFG sample, green circles are for the $z\sim0$ COLD~GASS SFG.
 Red colors are used for extreme galaxies such as the z$>$1 SMGs (star symbols), the $z\sim0$ COLD~GASS `major mergers' (squares) and $z\sim0$ ULIRGs (triangles).
 The standard deviations for the $z\sim0$ and $z\sim1$ samples include only the SFGs.}
 \label{fg:Mgas_SA_vs_CO}
\end{figure}

\begin{figure}[t]
 \centering
 \includegraphics[width=\columnwidth, bb=13 175 598 616]{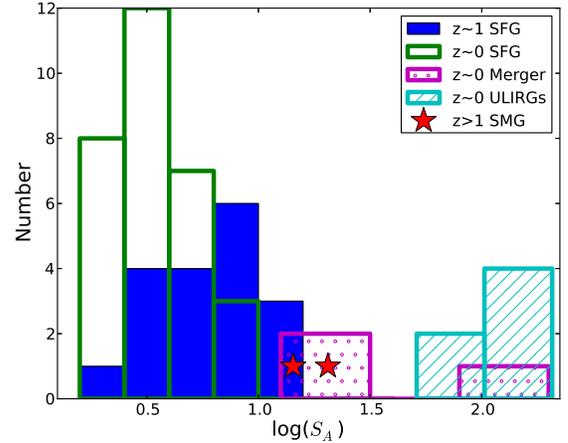}
 % S_A_histogram.eps: 0x0 pixel, 300dpi, 0.00x0.00 cm, bb=13 175 598 616
 \caption{Histograms of the $S_A$ values in the different samples. $S_A$ values are calculated from its definition using e.q.~\ref{eq:S_A_defintion}.}
 \label{fg:S_A_histogram}
\end{figure}

The relation obtained in figure~\ref{fg:specific attenuation} bottom (e.q.~\ref{eq:bestfit_SA}) combined with the definition of $S_A$ (e.q.~\ref{eq:S_A_defintion}), can be used to derive molecular gas masses from integrated UV and FIR fluxes:
\begin{equation}
\begin{array}{rl}
  \log\left( \frac{M_{mol}}{10^9\,M_\odot} \right) & = \\
  & \log\left( \frac{A_{\rm IRX} \cdot SFR}{M_\odot yr^{-1}} \right) -0.20(A_{\rm IRX} - 1.26\beta) +0.09 \\
\end{array}
\label{eq:Mmol from S_A}
\end{equation}
In figure~\ref{fg:Mgas_SA_vs_CO} we plot the predicted molecular gas mass from e.q.~\ref{eq:Mmol from S_A} versus the measured values from CO observations.
E.q.~\ref{eq:Mmol from S_A} gives a good prediction to the molecular gas mass for the galaxies in our $z\sim1$ sample, on which the relation was calibrated (blue circles).
The scatter is only 0.12~dex, with no bias.
However, it is much more interesting to test this method for estimating molecular gas masses on different samples that were not used in its calibration.
For this purpose we use a z$\approx$0 subsample from the COLD~GASS project\footnote{\url{http://www.mpa-garching.mpg.de/COLD\_GASS/}} \citep{Saintonge11a, Saintonge12}, local ultra luminous infrared galaxies (ULIRGs) and two $z>1$ SMGs in GOODS-N.

We select the COLD~GASS galaxies that have FUV and NUV fluxes from GALEX, and 60 \& 100 $\mu$m fluxes from IRAS.
This sample includes a few galaxies which are labeled as `mergers' that will be considered separately from the other star forming galaxies (SFGs).
The local ULIRGs are selected from the sample of \citet{Solomon97} cross-matched with the sample of \citet{Howell10} that provide GALEX fluxes. All the ULIRGs are treated as `major mergers'.
Full details regarding the ULIRGs, the SFG and COLD~GASS subsamples, as well as the derivation of IR luminosities and UV slopes are given in Appendix~\ref{app:COLDGASS_and_ULIRGs}.
A discussion regarding the conversion of IR luminosity to SFR in the COLD~GASS sample is provided in Appendix~\ref{app:LIR_discussion}.

We apply e.q.~\ref{eq:Mmol from S_A} on the COLD GASS subsample with no additional tuning.
The result is plotted in figure~\ref{fg:Mgas_SA_vs_CO}.
As we can see, e.q.~\ref{eq:Mmol from S_A} which was derived for $z=1$ galaxies, also produces good molecular gas mass estimates for the $z=0$ SFG (green circles), which are an order of magnitude lower in $M_{mol}$ and SFR than the $z=1$ galaxies.
The standard deviation is only 0.16~dex, slightly larger than the scatter at $z=1$, though on average, e.q.~\ref{eq:Mmol from S_A} tends to underestimate COLD~GASS $M_{mol}$ by 0.1~dex.
There are several possible causes to the 0.1~dex offset.
Changes to $\beta_0$, to the UV extinction curve, or to the dust-to-gas ratio can potentially explain a small shift in the $S_A$--$A_{\rm IRX}-\beta$ relation between $z=1$ and $z=0$, that will affect e.q.~\ref{eq:Mmol from S_A}.
While in the $z>1$ sample we avoided using filters that may suffer enhanced attenuation from a 2170~\AA\ `UV bump', we could not do so with GALEX. The NUV filter is very broad and overlaps with a possible UV bump often observed in local galaxies.
This will tend to lower $\beta$, which in turn leads to underestimation of $M_{mol}$.
In order to explain the 0.1~dex underestimation seen in Figure~\ref{fg:Mgas_SA_vs_CO}, $\beta$ will need to change by $\sim$0.4, meaning NUV lowered by $\sim$0.25~mag due to the `UV bump'.

`Major mergers' were not included in the $z=1$ sample on which the method was calibrated. The derived $M_{mol}$ for the $z=0$ `mergers' (COLD~GASS and ULIRGs, empty red symbols) are systematically over estimated, by an order of magnitude.
Two SMGs in GOODS-N have PACS fluxes and good optical (rest-UV) photometry: HDF169 and HDF242 at redshifts 1.22 and 2.49. The CO fluxes are from  \citet{Frayer08} and \citet{Ivison11} respectively. Being PACS sources in GOODS-N and at the suitable redshifts, these sources are also included in our \IRXbeta\ main sample (Section~\ref{sec:Data}).
The two SMGs are also likely to be products of major mergers which is also reflected in their $\alpha_{CO}\approx1$~$M_\odot$~K$^{-1}$~km$^{-1}$~s~pc$^{-2}$ conversion factors \citep{Tacconi08, Magdis11b, Magnelli12}, similar to the local ULIRGs.
The two SMGs are plotted in Figure~\ref{fg:Mgas_SA_vs_CO} as red stars.
Similar to the local mergers, the SMGs $M_{mol}$ are over estimated by the $S_A$ method, though only by $\sim$0.3~dex.

Galaxies that are undergoing major mergers appear to follow a different $S_A$--\IRXbeta\ relation than the tight relation we find for normal SFGs.
This likely reflects their compact, intense, central star forming region that dominates the total SFR.
If we compare the $S_A$ values (using arbitrary $\tau_{UV}=10^9$~yr~$M_\odot^{-1}$) we find a median of 6.4 and 3.4 $M_\odot^{-1}$ for the z=1 and z=0 SFGs, versus medians of 17, 40 and 134 $M_\odot^{-1}$ for the SMGs, COLD~GASS mergers and ULIRGs.
Figure~\ref{fg:S_A_histogram} shows the $S_A$ histograms of the various populations.
There is little to no overlap in the spread of $S_A$ between the SFGs and mergers.
The two SMGs in our sample seem to be very moderate examples and we expect other SMGs to reach higher $S_A$ values, more similar to local ULIRGs. 
The large difference in absolute $S_A$ in addition to the deviation from the $S_A$--\IRXbeta\ relation is indicative of a change in the typical star dust geometry, beyond a simple change in the compactness (local surface density) of the star forming regions.
For example, a transition from scattered star forming regions with variable obscuration to a central star forming region with dust and stars mix can lead to a dramatic increase in $S_A$.
We discuss this further in Section~\ref{sec:Discussion}.

E.q.~\ref{eq:Mmol from S_A} can therefore be used for molecular gas mass estimation on a wide selection of SFG galaxies.
We must caution that the method has been derived and tested on galaxies that are mostly SFG with $\beta>-1.5$.
Passive galaxies and/or very compact major mergers high above the main-sequence, may follow different $S_A$ scalings, as is the case for the COLD~GASS mergers and ULIRGs in our sample.
We further discuss the possible application to major mergers in the next section.

%%%%%%%%%%%%%%%%%%%%%%%%%%%%%%%%%%%%%%%%%%%%%%%%%%%%%%%%%
\subsection{Comparison With The Kennicut-Schmidt Relation}

\begin{figure}[t]
 \centering
 \includegraphics[width=\columnwidth]{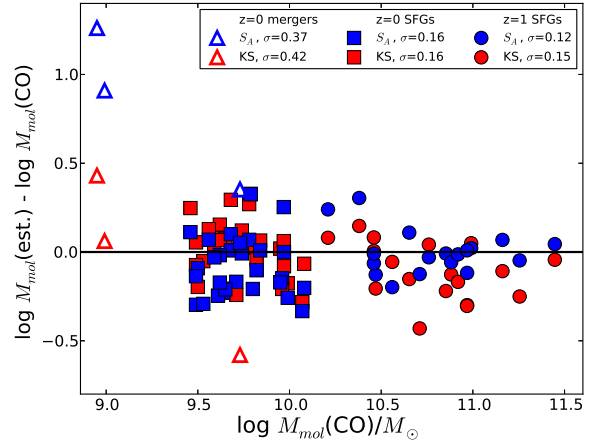}
 % CO_methods_compare.eps: 0s pixel, depict,  cm, BB=-55 175 667 616
 \caption{The deviation of the molecular mass, as derived by various methods, from the molecular mass as measured from CO.
	  Triangles are for assuming a constant gas depletion time scale, squares are for the inverse KS method and circles are for the $S_A$ method in this paper. The standard deviation for each method in each of the three samples is indicated in the legend.}
 \label{fg:CO_methods_compare}
\end{figure}

Another method commonly used for indirect estimation of gas mass is inverting the KS relation (inverse-KS).
In order to compare the $S_A$ method with the inverse-KS method, we use the three subsamples used above: The z$\sim$1 CO sample (Table~\ref{tab:CO sample}) and the z$\sim$0 FIR-bright COLD~GASS subsample (Table~\ref{tab:COLDGASS}) split into normal (labeled `SFGs')\footnote{Excluding COLD~GASS ID 3962 due to lack of a reliable radius measurement} and merger galaxies.
For the inverse-KS method, we adopt the relations of \citet[][]{Genzel10}:
\begin{equation}
\log\left( \frac{\Sigma_{SFR}}{M_\odot{\rm yr}^{-1}{\rm Kpc}^{-2}} \right) = 1.17 \log\left( \frac{\Sigma_{mol-gas}}{M_\odot{\rm pc}^{-2}} \right) - 3.48 
\label{eq:KS_for_SFG}
\end{equation}
\begin{equation}
\log\left( \frac{\Sigma_{SFR}}{M_\odot{\rm yr}^{-1}{\rm Kpc}^{-2}} \right) = 1.17 \log\left( \frac{\Sigma_{mol-gas}}{M_\odot{\rm pc}^{-2}} \right) - 2.44 
\label{eq:KS_for_mergers}
\end{equation}
were e.q.~\ref{eq:KS_for_SFG} is used for the normal galaxies and e.q.~\ref{eq:KS_for_mergers} for the major mergers.
For the galaxies surface area we adopt the half light radii from the corresponding literature sources of the samples.
The half SFRs are calculated from the $SFR({\rm IR})+SFR({\rm UV})$ as described in Section~\ref{sec:terms_and_definitions}.

In Figure~~\ref{fg:CO_methods_compare} we plot the deviation of the estimated $M_{mol}$ from the value derived from CO measurements in dex.
The standard deviation (dex) for each subsample with each method is displayed in the legend.
The three subsamples are treated separately so possible systematic offsets between the samples will not translate into an increase in the standard deviation.
Biases can be adjusted by calibration, but the scatter is more inherent to the method.
For the samples used in this study the $S_A$ method gives molecular gas mass estimates with an accuracy as good or better than the inverse-KS method, when avoiding extreme cases of major mergers.
This is achieved using only integrated rest frame UV and FIR photometry without the need to resolve the galaxies.

For 'major merger' galaxies, the $S_A$ method has a large systematic bias in the derived molecular masses versus the inverse-KS method.
The latter treats SFG and major mergers separately, using two different relations.
However, for the local mergers, the scatter around the systematic bias of the $S_A$ method is not larger (slightly smaller in our sample) than the scatter in the inverse-KS relation, which suggests that it may be possible to derive a separate calibration for major mergers, that removes the bias.
For example, in our sample, lowering the mergers $M_{mol}$ by 0.8~dex will place the $S_A$ mass estimates nearly on top of the inverse-KS estimates, object by object.
0.8~dex is the difference between using e.q.~\ref{eq:KS_for_SFG} and e.q.~\ref{eq:KS_for_mergers} for these galaxies.
Since here the 'major mergers' include only 3 objects we cannot come to any solid conclusion regarding the application of the $S_A$ method to such galaxies. A much larger sample will be required in order to do so.

%%%%%%%%%%%%%%%%%%%%%%%%%%%%%%%%%%%%%%%%%%%%%%%%%%%%%%%%%%%%%%%
%
\section{Discussion: Dust Obscuration and Geometry} 
\label{sec:Discussion}
%
%%%%%%%%%%%%%%%%%%%%%%%%%%%%%%%%%%%%%%%%%%%%%%%%%%%%%%%%%%%%%%%

\subsection{Effective Attenuation and The Attenuation Distribution} 
\label{sec:attenuation_distribution}

In the previous sections we found that the scatter in the \IRXbeta\ plane correlate with the distance from the main sequence (Figure~\ref{fg:A-beta dMS colors}) and also with the specific attenuation $S_A$ (Figure~\ref{fg:specific attenuation}).
What is the cause of the scatter and why does $S_A$ correlates so well with $A_{\rm IRX}-A_{\rm MS}$?
The physical parameters involved are the dust properties (extinction curve), the stellar population (unattenuated spectral slope) and the geometrical arrangement of the dust and stars.
We must also keep in mind that here we do not observe a single star-forming region, but instead integrate the light from the entire galaxy.
Hence, $A_{\rm IRX}$ is an {\it effective} UV attenuation that may differ from the attenuation of each individual source in the galaxy.
To better understand this, we examine in this section the effects of integration on the $A_{\rm IRX}$--$\beta$ relation.
Full derivations for the equations below are given in Appendix~\ref{app:analytical_IRX_beta}.

When integrating the luminosity over an entire galaxy, we include many individual point-like sources and each may have a different attenuation value.
Let us denote the normalized distribution of attenuations of the individual sources in the galaxy as $f_{\lambda_1}(A)$, i.e. the fraction of sources suffering attenuation between $A$ and $A+{\mathrm d}A$ at wavelength $\lambda_1$.
Let us now consider a general representation for $f_{\lambda_1}$.
We will assume that most of the UV sources are scattered inside a typical range of attenuation values between $A_{min}$ and $A_{min} + {\Delta}A$.
For simplicity, we will assume that the distribution is flat inside this range.
\begin{equation}
  f_{\lambda_1}(A) = \left\{ \begin{array}{l l}
                    1/\Delta A &\quad A_{min}<A<A_{min}+\Delta A \\
		    0            &\quad else \\
                   \end{array} \right.
\label{eq:slabmix_dist}
\end{equation}
Due to the integration over all sources, the resulting effective attenuation is not very sensitive to the fine details of $f_{\lambda_1}$, with a possible exception around $A \to 0$. Most distributions that are unimodal can be approximated as $f_{\lambda_1}$ above.

We can now obtain the solution for effective attenuation at wavelength $\lambda_1$:
\begin{equation}
 A_{\lambda_1} = A_{min} -2.5 \log \left( \frac{ \log(e) \left( 1-10^{-0.4{\Delta}A} \right) }{ 0.4 {\Delta}A } \right)
\label{eq:A_for_slabmix}
\end{equation}
By our definitions and when using the same rest UV wavelength in $\lambda_1$ and for deriving $A_{\rm IRX}$, $A_{\lambda_1}$ and $A_{\rm IRX}$ are equivalent.
The observed spectral slope $\beta$ is then given by:
\begin{equation}
 \beta - \beta_0 = \frac{1}{\log(R_{1,2})} \left[ -0.4 \left( 1-R_{1,2}^{-\gamma} \right) A_{min} +  \log\left( \frac{1-10^{-0.4{\Delta}A}}{1-10^{-0.4{\Delta}A R_{1,2}^{-\gamma}}} \right) \right]
\label{eq:beta_for_slabmix}
\end{equation}
where $\gamma$ is the power of the UV attenuation law ($A(\lambda)\propto\lambda^\gamma$) and $R_{1,2} = \lambda_1/\lambda_2$. $\lambda_1$ and $\lambda_2$ are the two wavelengths between which the UV slope is measured, 1600~\AA\ and 2800~\AA\ respectively in this study.
With equations~\ref{eq:A_for_slabmix} and \ref{eq:beta_for_slabmix} above we can plot the expected $A_{\rm IRX}$--$\beta$ relation for various $f_{\lambda_1}(A)$ distributions.

\begin{figure}[t]
 \centering
 \includegraphics[width=\columnwidth]{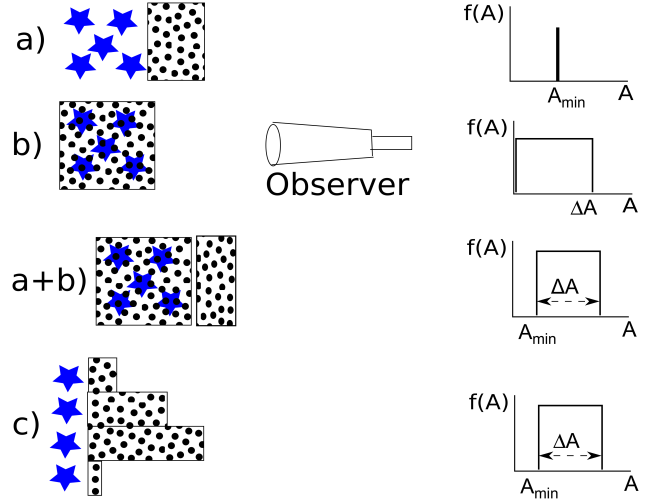}
 %\includegraphics[width=\columnwidth, clip=true, trim=capt 0pt 0pt adapt]{./obscuration.eps}
 % obscuration.eps: 0s pixel, depict,  cm, BB=-86 0 467 738
 \caption{A sketch of three basic geometrical arrangements for the gas/dust and stars. 
{\it a)} The obscuring screen, {\it b)} Dust and stars mix, {\it c)} variable screens.
{\it a$+$b}~is a combination of a mix with an obscuring screen.
The attenuation distribution for each case is plotted on the right hand side and parametrized by $A_{min}$ and $\Delta A$}
 \label{fg:obscurations sketch}
\end{figure}

\subsection{Archetypal Dust and Star Geometries}
\label{sec:Archtypal_geometries}

\begin{figure}[t]
 \centering
 \includegraphics[width=\columnwidth]{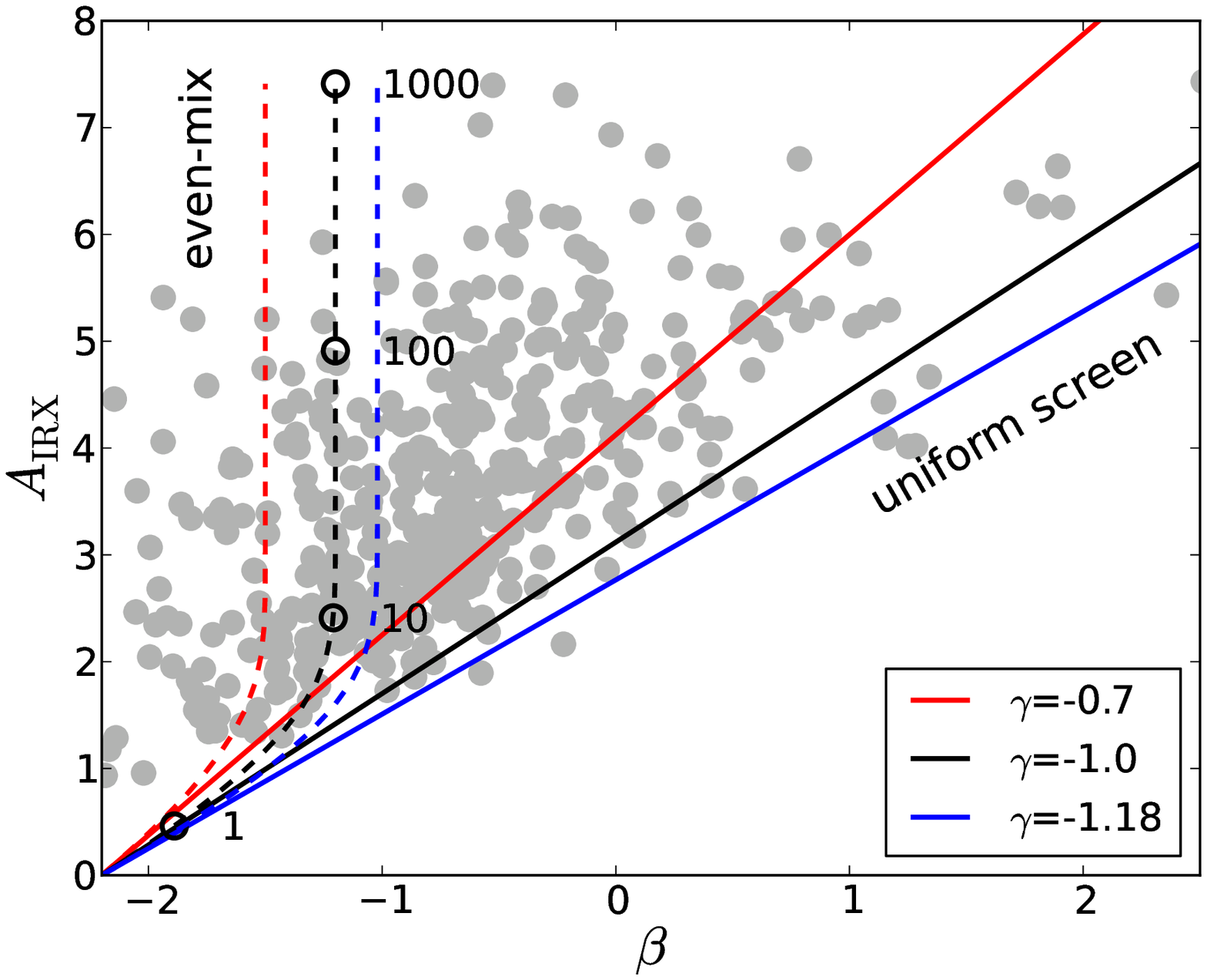} \\
 \includegraphics[width=\columnwidth]{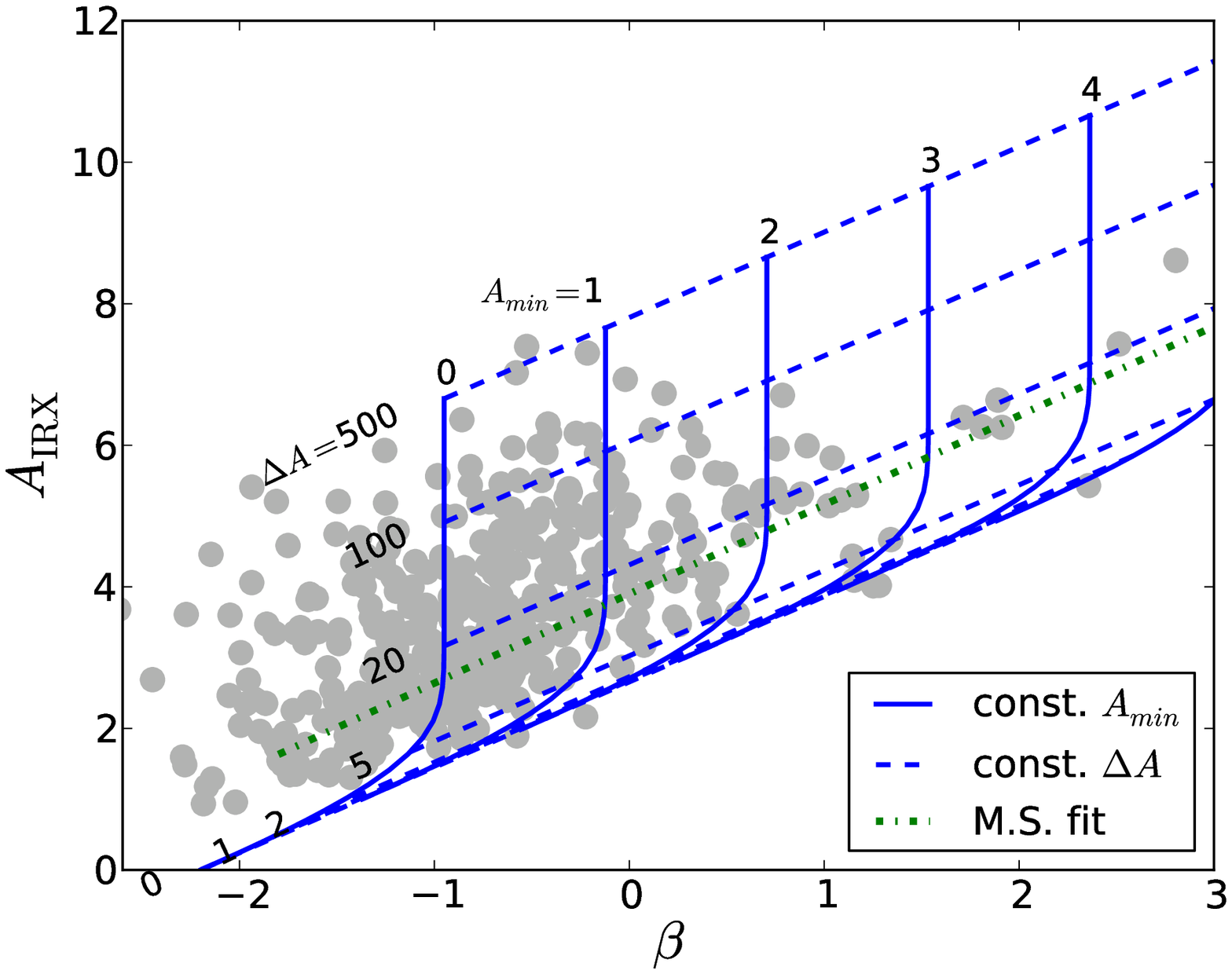}
 % A_beta_geometry_models.eps: 0s pixel, depict,  cm, BB=13 175 598 616
 \caption{{\it Top:} $A_{\rm IRX}$--$\beta$ relations for a uniform slab model (solid lines) and even-mix (dashed lines). The colors are for various values of the UV extinction law slope $\gamma$. The values along the dashed line indicate the $A_{max}$ value for the even-mix model.
 The data is plotted in gray in the background.
  {\it Bottom:} $A_{\rm IRX}$--$\beta$ relations for various values of a flat attenuation distribution starting at $A_{min}$ with width $\Delta A$. Lines of constant $A_{min}$ are plotted in solid red. Lines of constant $\Delta A$ are plotted in dashed red. The relation for main-sequence galaxies is plotted in dash-dot green. The data is plotted in gray in the background.}
 \label{fg:beta_for_even_mix}
\end{figure}

Let us now consider some possible, basic arrangements for the stars and dust mix.
In figure~\ref{fg:obscurations sketch} we sketch three archetypal stars and dust arrangements and one combination.
These are by no means all the possible arrangements, but are sufficient for demonstrating the general principle.
On the right hand side of the figure, we plot the approximation for the resulting $f_{\lambda_1}(A)$.
For simplicity, but still maintaining the essence of the matter, in all cases $f_{\lambda_1}$ is approximated to be a flat distribution between $A_{min}$ and $A_{min} + {\Delta}A$.

The two most commonly assumed geometries are the uniform obscuring screen (case~{\it a}) and the even-mix (case~{\it b}) of stars and dust.
The $A_{\rm IRX}$--$\beta$ relation for both cases can be obtained from the general solution (e.q.~\ref{eq:A_for_slabmix} and \ref{eq:beta_for_slabmix}).
The special solutions for each case can be found in Appendix~\ref{app:relation_for_archtypal_cases}.

In Figure~\ref{fg:beta_for_even_mix} top panel we plot the expected $A_{\rm IRX}$--$\beta$ relations for uniform screen and even-mix models with various attenuation laws, while fixing $\beta_0=-2.2$.
For the uniform obscuring screen, the \IRXbeta\ relation is linear with $\beta$ (e.q.~\ref{eq:A_IRX_slab}).
The slope is determined solely by $\gamma$, while $\beta_0$ determines the intercept with the $\beta$ axis.
Variation in the SSFR will change the ratio between the young and old stellar populations that tend to have lower and higher $\beta_0$ respectively, resulting in a horizontal shift in the $A_{\rm IRX}$--$\beta$ plot.
Variations in $\beta_0$ can thus create scatter in the $A_{\rm IRX}$--$\beta$ relation that is correlated with the SSFR and $\Delta\log({\rm SFR})_{\rm MS}$.
However, even main-sequence galaxies at $z>1$ have very high SSFRs, comparable with intense local starbursts and their UV light is still dominated by the very young population.
To explain the large scatter in $A_{\rm IRX}$ by $\beta_0$ variations alone (under the uniform screen assumption) requires very low $\beta$ values, lower than those seen in young stellar clusters and outside the typical predicted range of $-2.6<\beta<-2.0$ \citep[e.g., ][]{Wilkins12}.
The relation that we derived from the main sequence galaxies in section~\ref{sec:A vs beta} implies $\beta_0=-3.0$, that is an unusually low value.
However, the direct translation of the intercept in the linear $A_{\rm IRX}$--$\beta$ relation into $\beta_0$ is relevant only if the geometry is that of a uniform obscuring screen, which is disfavored in this study.

The even-mix geometry (Figure~~\ref{fg:beta_for_even_mix}, dashed lines) also fails to explain the data.
The even-mix distribution can only produce very blue (i.e. low) $\beta$.
The limit for the change in the UV spectral slope is:
\begin{equation}
 \lim_{\Delta A \to \infty} \beta-\beta_0 = -\gamma
\end{equation}
With typical values of $\gamma \sim -1$, this distribution can only produce a narrow range of low $\beta$ values, lower than in the majority of our sample.
It is clear therefore, that a broad $f_{\lambda_1}(A)$ distribution that starts from near zero attenuation does not fit our sample.
The typical galaxies in the sample must have $f_{\lambda_1}(A)$ distributions that vanish when $A \to 0$ and that effectively means at least some foreground screen component. In other words, $A_{min} > 0$ and $\Delta A > 0$ are required.

We denote the combination of even-mix with an additional foreground screen as case~{\it a+b}.
Such a distribution could arise from young stars embedded in molecular clouds starting at a minimal depth, or an additional obscuring component in the diffuse interstellar medium (ISM).
Perhaps just as relevant when dealing with integrated luminosities, is the variable screens geometry as sketched in figure~\ref{fg:obscurations sketch}~{\it c}.
In case {\it c}, each star forming region has a different screen and the region-to-region variations are what produces the $\Delta A$ width of $f_{\lambda_1}(A)$, as opposed to case~{\it a+b} where $\Delta A$ is the depth inside a slab.
As long as the resulting $f_{\lambda_1}(A)$ in cases {\it c} and {\it a+b} are similar, they will produce a similar \IRXbeta\ relation.
%The distinction is more than just semantics when it comes to $S_A$ (Section~\ref{sec:S_A_geometry_discussion}).
Adopting $\gamma=-1.18$ from $A_{\rm MS}$ (e.q.~\ref{eq:A_MS}), $\beta_0=-2.2$ \citep[between ][]{Meurer99, Calzetti00}, and variable $A_{min}$ and $\Delta A$ values, we plot the resulting $A_{\rm IRX}$--$\beta$ relations in figure~\ref{fg:beta_for_even_mix} bottom panel.
In this picture, $z\sim1$ main-sequence galaxies have a typical attenuation distributions with full widths of $\Delta A \sim 12$ magnitudes, while galaxies with enhanced SFR have much broader distributions.
The main sequence galaxies scatter along the linear relation mostly due to the $A_{min}$ component that varies between them and controls the observed UV color.
Being the most general case that combines both a screen and a broad distribution we can see in Figure~\ref{fg:beta_for_even_mix} that no $A_{min}$ and $\Delta A$ combination can explain galaxies with $\beta \lesssim -1.5$ and $A_{\rm IRX}>1$.
These galaxies either have a combination of lower $\beta_0$ and a flatter attenuation law (higher $\gamma$), or more likely, a bimodal-like attenuation distributions (Appendix~\ref{app:bimodal_distribution}).

\subsection{The Relation Between $S_A$ and Geometry}
\label{sec:S_A_geometry_discussion}

On the $A_{\rm IRX}$--$\beta$ plane there is no difference between a combination of stars--dust mix with an additional obscuring screen (case~{\it a}+{\it b}) and the variable screens (case~{\it c}), as long as the resulting attenuation distributions $f_{\lambda_1}(A)$ are similar.
There is however an important distinction between the $S_A$ of the two cases: in case {\it a+b}, each layer of the slab obscure several stars, the observer-side layer obscuring the largest number. This allows a relatively small amount of gas to provide a high attenuation for most stars, as opposed to case~{\it c}, where each star/region requires its own independent column of gas, resulting in a lower $S_A$ (i.e., more gas per UV star to achieve the same integrated $A_{\rm IRX}$).

As we can see from e.q.~\ref{eq:A_for_slabmix}, if the attenuation distribution is unimodal, then the deviation of the galaxy from the uniform screen (for which $A_{\rm screen}=A_{min}$) \IRXbeta\ relation behaves like $\sim \log(\Delta A)$, regardless if $\Delta A$ is an actual slab depth (case~{\it b}) or a variation between regions (case~{\it c}).
The correlation of $S_A$ with $A_{\rm IRX} - A_{\rm screen}(\beta)$ is caused by the width of the $f_{\lambda_1}(A)$ distribution, represented in our simplified model by $\Delta A$, and regardless of $A_{min}$.
An increase in $\Delta A$ move the galaxies away from $A_{\rm screen}(\beta)$ and corresponds to a more compact star forming regions or a stronger tendency towards a mix of stars and gas, thus $S_A$ increases as well. 
Note that the values of $A_{min}$ are significantly smaller than $\Delta A$, and for most of our sample, $A_{min}<2$.
This means that most of the gas mass is in the $\Delta A$ component and only a small fraction of the gas is in the foreground.
This is likely the reason why $A_{\rm IRX}-A_{\rm screen}(\beta)$, that is nearly proportional to $\log(\Delta A)$, is such a good indicator to the gas mass in e.q.~\ref{eq:Mmol from S_A}.

The galaxies from which e.q.~\ref{eq:bestfit_SA} was derived are all star forming disks, dotted with star forming regions and likely share a fairly similar basic stars--dust geometry.
The same can be said about the COLD~GASS SFGs.
Moderate changes to the surface densities and/or geometry are enough to explain the limited range of $S_A$ in these two SFGs samples (Figure~\ref{fg:S_A_histogram}).
The local ULIRGs on the other hand are characterized by a central core of very dense star formation. The ULIRGs are also likely to have a fundamentally different obscuration geometry, that results in their $S_A$ being higher by two orders of magnitude than that of the SFGs.
In merger galaxies $S_A$ increase both by the compactness of the star forming region as well as due to a radical change in the geometry type, thus they deviate from the relation found for SFGs (e.q.~\ref{eq:bestfit_SA} and e.q.~\ref{eq:Mmol from S_A}).
Mergers are also more likely to have a bimodal $f_{\lambda_1}(A)$, which will throw off the relation between $S_A$ and $\beta$.

%%%%%%%%%%%%%%%%%%%%%%%%%%%%%%%%%%%%
%
\section{Conclusions} \label{sec:conclusions}
%
%%%%%%%%%%%%%%%%%%%%%%%%%%%%%%%%%%%%

In this work we studied the attenuation of UV (1600~\AA) light from star forming galaxies at redshifts $1<z<2.5$ and its relation to the UV slope $\beta$.
The effective attenuation $A_{\rm IRX}$ was derived from the FIR/UV ratio, also known as IRX. 
We use {\it Herschel}-PACS deep FIR imaging and optical photometry from various instruments that cover rest-frame UV.
Our sample shows a significant scatter in the $A_{\rm IRX}$--$\beta$ relation, but with a structure to the scatter,
where galaxies with increasingly enhanced SFR relative to the main SFR--$M_*$ relation (main-sequence) show an enhanced $A_{\rm IRX}$ for their $\beta$.

When selecting only main-sequence galaxies, they tend to form a tighter sequence than the full population in the $A_{\rm IRX}$--$\beta$ diagram, with a flatter slope than commonly used relations, such as \citet{Meurer99} and \citet{Calzetti00}.
In terms of the UV extinction curve, our findings favor a steeper $A(\lambda) \propto \lambda^\gamma$ dependency than the above mentioned relations, with a power of $\gamma \sim -1.18$, similar to that of the Small Magellanic Cloud. However, such a conversion from the $A_{\rm IRX}$--$\beta$ slope to the extinction curve is geometry dependent.
The above mentioned literature $A_{\rm IRX}$--$\beta$ relations and the relation for main sequence galaxies found in this work cross each other close to the peak in the $\beta$ number distribution of $1<z<2.5$ galaxies. Hence, the attenuation corrections obtained from the different relations tend to produce similar global SFRs, but with different biases.

Using a smaller sample of $1<z<1.5$ galaxies with CO gas mass measurements, we studied the relation between UV attenuation and molecular gas masses.
We find a very tight relation between the location in the $A_{\rm IRX}$--$\beta$ diagram and $S_A$ the `specific attenuation', a quantity that represents the attenuation contributed by the amount of molecular gas available per young star.
$S_A$ depends on both the geometrical stars--dust arrangement and the compactness of the {\it individual}\footnote{Not to be confused with the `compactness' of the entire galaxy as expressed by the SFR surface density, though the two quantities may be correlated.} star forming regions.

We demonstrate that the tight empirical relation between $S_A$ and the location in the $A_{\rm IRX}$--$\beta$ diagram can be used to estimate molecular gas masses using UV and FIR photometry alone, without the need to resolve the galaxies.
For normal SFG at $0<z<1.5$, the accuracy is between 0.12--0.16~dex when compared with masses derived from CO observations.
Major mergers and SMGs seem to have a different stars and dust geometry and possibly follow a different $S_A$--\IRXbeta\ relation, that remains to be determined.

%%%%%%%%%%%%%%%%%%%%%%%%%%%%%%%%%%%%%%%%%%%%%%%%%%
%%%%%%%%%%%%%%%%%%%%%%%%%%%%%%%%%%%%%%%%%%%%%%%%%%
\acknowledgements
PACS has been developed by a
consortium of institutes led by MPE
(Germany) and including UVIE (Austria); KUL, CSL, IMEC (Belgium); CEA,
OAMP (France); MPIA (Germany); IFSI, OAP/OAT, OAA/CAISMI, LENS, SISSA
(Italy); IAC (Spain). This development has been supported by the funding
agencies BMVIT (Austria), ESA-PRODEX (Belgium), CEA/CNES (France),
DLR (Germany), ASI (Italy), and CICYT/MCYT (Spain).

This study makes use of data from AEGIS, a multi wavelength sky survey conducted with the Chandra, GALEX, Hubble, Keck, CFHT, MMT, Subaru, Palomar, Spitzer, VLA, and other telescopes and supported in part by the NSF, NASA, and the STFC.

%%%%%%%%%%%%%%%%%%%%%%%%%%%%%%%%%%%%%%%%%%%%
\bibliographystyle{apj}
\bibliography{bibli}
%%%%%%%%%%%%%%%%%%%%%%%%%%%%%%%%%%%%%%%%%%%%%
% latex x bibtex x latex x latex x

%%%%%%%%%%%%%%%%%%%%%%%%%
%  FIGURES   %%
%%%%%%%%%%%%%%%%%%%%%%%%%

%%%%%%%%%%%%%%%%%%%%%%%%%%%%%%%%%%%%%%%%%%%%%%
%
\appendix
%
%%%%%%%%%%%%%%%%%%%%%%%%%%%%%%%%%%%%%%%%%%%%%%

%%%%%%%%%%%%%%%%%%%%%%%%%%%%%%%%%%%%%%
\section{A. The $\chi_\perp^2$ Test}
\label{app:chisqr_test}
%%%%%%%%%%%%%%%%%%%%%%%%%%%%%%%%%%%%%%
This test uses a Monte-Carlo simulation in order to obtain the distribution of values which are derived from several observables with random errors.
Both axes in Figure~\ref{fg:specific attenuation} are derived values with propagated errors and therefore must be treated equally.
In addition, several observables are shared between the axes and therefore the propagated errors are correlated.
By calculating $\chi^2$ for the perpendicular deviation from a theoretical linear relation one can determine the `goodness of fit' without bias towards any of the two variables (axes).
By simulating a resampling of the observed data one can measure the distribution of the derived values from the simulation and do so simultaneously for both axes, deriving the distribution of the perpendicular distance from the tested theory. 
Thus, the correlations in the errors are also taken into account.

We randomize all the input fluxes (UV, FIR, CO) according to their estimated errors while assuming a normal distribution.
From the new set of fluxes we recalculate $\tau_{dep}$, $S_A$, $A_{\rm MS}(\beta)$ and $A_{\rm IRX}$. We then calculate the perpendicular distance $\delta_{\perp,i}$ from the original relations (in Figure~\ref{fg:specific attenuation}) for each galaxy $i$.
The process is repeated a large number of times until we can derive a distribution for each $\delta_{\perp,i}$.
The $\chi_\perp^2$ is then defined as:
\begin{equation}
 \chi_\perp^2 = \displaystyle\sum_i{ \frac{\hat{\delta}_{\perp,i}^2}{\sigma(\delta_{\perp,i})^2} }
\end{equation}
where $\hat{\delta}_{\perp,i}$ is the perpendicular distance of the measured data point $i$ from the mean relation and $\sigma(\delta_{\perp,i})$ is the standard deviation in $\delta_{\perp,i}$ as derived from the simulated distribution.
For each simulated set of input fluxes we also re-fit the $S_A(A_{\rm IRX}-A_{\rm MS})$ relation.
The distributions in the values of the coefficients are fully consistent with the best fit and uncertainty values given in e.q.~\ref{eq:bestfit_SA}.
%The $\chi_\perp^2$ values are indicated in each panel and indeed $S_A$ offers an improvement in tightening the relation.

%%%%%%%%%%%%%%%%%%%%%%%%%%%%%%%%%%%%%%%%%%%%%%%%%%%%
\section{B. The COLD GASS Subsample and Local ULIRGs}
\label{app:COLDGASS_and_ULIRGs}
%%%%%%%%%%%%%%%%%%%%%%%%%%%%%%%%%%%%%%%%%%%%%%%%%%%%

We select the COLD~GASS galaxies that have FUV and NUV fluxes from GALEX, and 60 \& 100 $\mu$m fluxes from IRAS.
Galaxy pairs, whose individual components are confused in some of the bands were excluded.
$\beta$ and the UV flux at 1600~\AA\ were calculated by fitting a power-law spectrum to the GALEX fluxes, while taking into account the (very broad) filters responses.
The $L_{IR}(8-1000\,\mu{\rm m})$ luminosity was calculated by fitting CE01 templates to the two IRAS fluxes, while allowing both scale and template to vary.
For a discussion on the conversion of $L_{IR}$ to SFR for this sample see Appendix~\ref{app:LIR_discussion}.
Most of these galaxies tend to be SFG slightly above the $z=0$ main sequence \citep[e.g., ][]{Brinchmann04} and we adopt $\alpha_{CO}=4.35$~$M_\odot$~K$^{-1}$~km$^{-1}$~s~pc$^{-2}$ conversion factor for these galaxies (similar to the $z\sim1$ sample).
A few of the selected galaxies are classified as `mergers' and a different conversion factor ($\alpha_{CO}=1.0$~$M_\odot$~K$^{-1}$~km$^{-1}$~s~pc$^{-2}$) was used in the derivation of their $M_{mol}$.
The criteria for using the lower $\alpha_{CO}$ are IRAS flux ratio S(60)/S(100)$>$0.6 and $L_{\rm IR}>10^{11}$~$L\odot$.
Two of the three galaxies flagged as `mergers' also show clear signs of an ongoing merger in their optical images.
The essential information of the COLD GASS subsample is displayed in Table~\ref{tab:COLDGASS}.

In order to include more extreme local galaxies, we build a sample of ULIRGs composed of objects shared between the samples of \citet{Solomon97} and \citet{Howell10}.
These ULIRGs were observed with GALEX and IRAS and we treat them in the same way as the COLD~GASS sample. 
\citet{Solomon97} found for these galaxies $\alpha_{CO} \approx 1.0$ and we adopt this as a uniform value for all ULIRGs in the sample.
See Table~\ref{tab:Solomon_Howell} for the list of objects and photometric fluxes.

\begin{figure}[h]
 \centering
 \includegraphics[width=0.5\columnwidth, bb=13 175 598 616]{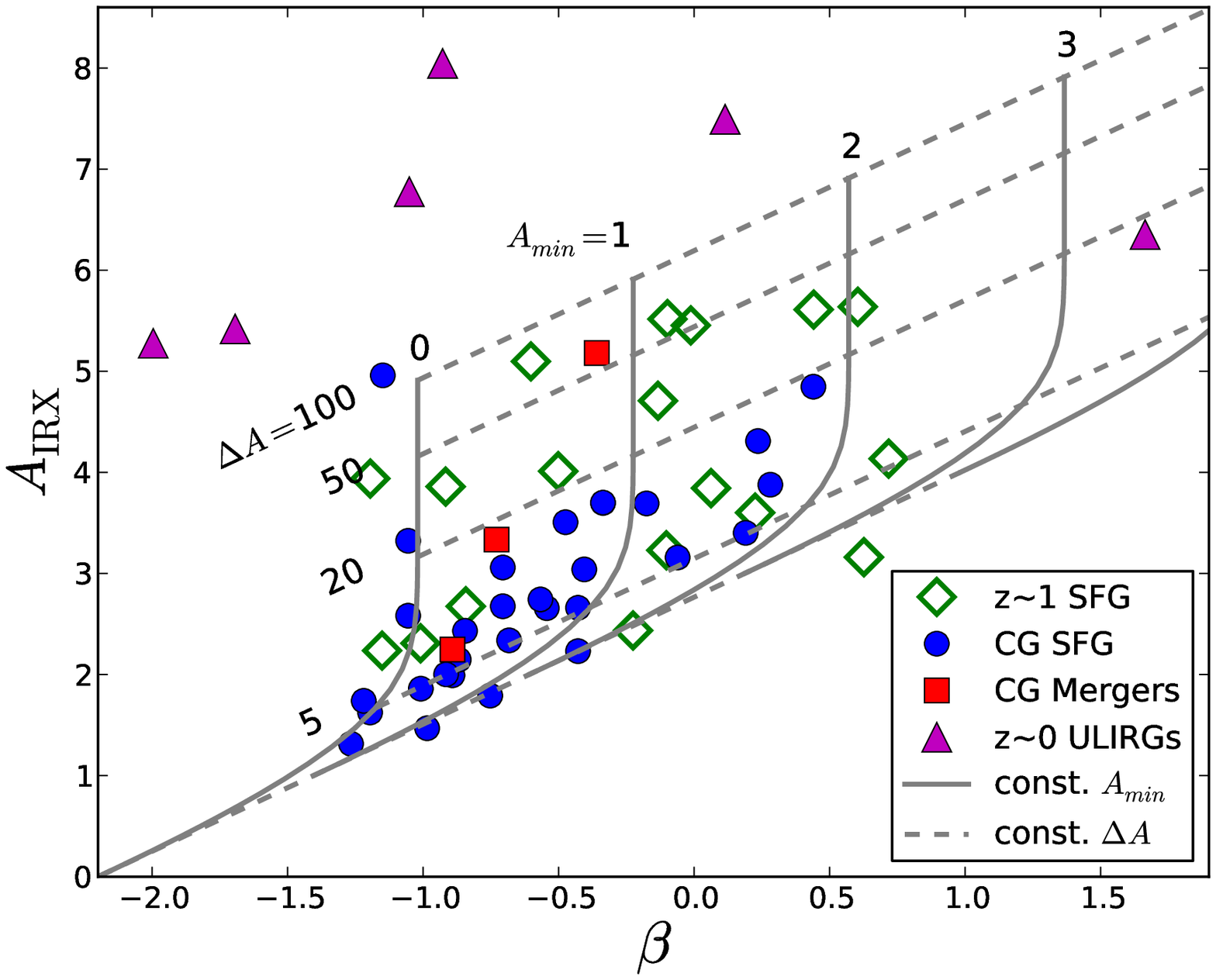}
 % Local_sample_IRXbeta.eps: 0x0 pixel, 300dpi, 0.00x0.00 cm, bb=13 175 598 616
 \caption{\IRXbeta\ relation for the local samples: COLD~GASS (CG) SFGs in blue circles, COLD~GASS `mergers' in red squares, local ULIRGs in magenta triangles. The z$\sim$1 CO sample is plotted in empty green diamonds for reference. The gray grid indicate lines of constant $A_{min}$ and $\Delta A$, identical to Figure~\ref{fg:beta_for_even_mix}}
 \label{fg:Local_IRX_beta}
\end{figure}

In Figure~\ref{fg:Local_IRX_beta} we plot the location of the COLD~GASS and ULIRGs samples on the \IRXbeta\ plane. The z$\sim$1 sample is also plotted for comparison. Also plotted is the grid of \IRXbeta\ relations for the various attenuation distributions, similar to Figure~\ref{fg:beta_for_even_mix} (see Section~\ref{sec:Archtypal_geometries} for details).

\begin{table*}[h]
\begin{center}
  \caption{\label{tab:COLDGASS} The COLD GASS subsample \citep{Saintonge11a} with GALEX and IRAS detections}
  \begin{tabular}{l*{10}{c}}
    ID & z & NUV & FUV & $\beta$ & S60 & S100 & $L_{IR}$           & $A_{\rm IRX}$ & $\log M_{mol}^\dagger$ & $\alpha_{\rm CO}^\ddagger$ \\
       &   & AB  & AB  &         & Jy  &  Jy  & 10$^{10}$ L$_\odot$& mag           & M$_\odot$ & \\ 
    \hline\hline
3819 & 0.0453 & 17.27 & 17.89 & -0.54 & 0.78 & 1.44 & 11.05 & 2.66 & 9.97 & 4.35\\
3962 & 0.0427 & 18.38 & 19.09 & -0.34 & 0.57 & 1.39 & 9.02 & 3.7 & 10.06 & 4.35\\
20286 & 0.0346 & 20.35 & 20.71 & -1.15 & 0.34 & 0.99 & 4.16 & 4.96 & 9.49 & 4.35\\
26221 & 0.0317 & 18.03 & 18.81 & -0.18 & 0.58 & 1.8 & 6.34 & 3.69 & 10.07 & 4.35\\
29842 & 0.0341 & 18.15 & 19.09 & 0.19 & 0.27 & 1.09 & 4.37 & 3.4 & 9.68 & 4.35\\
24741 & 0.0492 & 17.64 & 18.11 & -0.89 & 0.25 & 0.67 & 5.26 & 1.99 & 9.62 & 4.35\\
6375 & 0.0488 & 17.46 & 17.95 & -0.85 & 0.38 & 1.11 & 9.57 & 2.43 & 9.78 & 4.35\\
6506 & 0.0487 & 18.1 & 18.71 & -0.57 & 0.32 & 0.85 & 6.57 & 2.74 & 9.99 & 4.35\\
40781 & 0.048 & 18.82 & 19.22 & -1.06 & 0.27 & 0.82 & 6.89 & 3.32 & 9.8 & 4.35\\
7493 & 0.0264 & 18.41 & 19.12 & -0.34 & 0.64 & 1.53 & 3.28 & 3.7 & 9.64 & 4.35\\
10019 & 0.0308 & 17.67 & 18.15 & -0.87 & 0.19 & 0.7 & 2.29 & 2.15 & 9.53 & 4.35\\
41969 & 0.0351 & 16.59 & 16.9 & -1.27 & 0.38 & 0.93 & 3.52 & 1.31 & 9.61 & 4.35\\
4216 & 0.0438 & 17.24 & 17.64 & -1.06 & 0.77 & 1.96 & 11.76 & 2.58 & 10.08 & 4.35\\
42013 & 0.0368 & 19.2 & 20.25 & 0.44 & 0.75 & 1.39 & 6.94 & 4.85 & 9.96 & 4.35\\
4045 & 0.0265 & 17.52 & 18.2 & -0.41 & 1.01 & 1.68 & 4.07 & 3.04 & 9.71 & 4.35\\
19949 & 0.0289 & 18.67 & 19.63 & 0.24 & 0.3 & 1.61 & 4.49 & 4.31 & 9.56 & 4.35\\
56632 & 0.0281 & 18.83 & 19.81 & 0.28 & 0.23 & 0.89 & 2.4 & 3.88 & 9.5 & 4.35\\
14712 & 0.0383 & 17.14 & 17.81 & -0.43 & 0.32 & 1.05 & 5.45 & 2.23 & 9.82 & 4.35\\
18673 & 0.0385 & 18.69 & 19.34 & -0.47 & 0.38 & 1.03 & 4.79 & 3.51 & 9.59 & 4.35\\
50550 & 0.035 & 18.56 & 19.11 & -0.71 & 0.21 & 0.73 & 3.13 & 3.06 & 9.49 & 4.35\\
28365 & 0.0321 & 16.61 & 16.95 & -1.2 & 0.34 & 1.15 & 4.12 & 1.62 & 9.65 & 4.35\\
31775 & 0.0412 & 16.68 & 17.23 & -0.71 & 1.38 & 2.71 & 16.87 & 2.68 & 9.79 & 4.35\\
33039 & 0.0434 & 16.56 & 16.89 & -1.22 & 0.67 & 1.34 & 9.26 & 1.74 & 9.68 & 4.35\\
12533 & 0.0331 & 16.64 & 17.17 & -0.75 & 0.35 & 1.16 & 4.44 & 1.79 & 9.46 & 4.35\\
30332 & 0.0425 & 16.7 & 17.12 & -1.01 & 0.6 & 1.27 & 8.3 & 1.86 & 9.95 & 4.35\\
13005 & 0.0487 & 17.4 & 17.96 & -0.68 & 0.52 & 0.97 & 8.63 & 2.34 & 9.84 & 4.35\\
25327 & 0.0314 & 16.73 & 17.27 & -0.73 & 3.35 & 5.07 & 17.8 & 3.34 & 9.73 & 1.0\\
13666 & 0.0405 & 16.72 & 17.19 & -0.89 & 1.2 & 1.84 & 10.79 & 2.25 & 8.99 & 1.0\\
44942 & 0.0329 & 16.42 & 16.88 & -0.92 & 0.89 & 1.86 & 7.2 & 2.0 & 9.74 & 4.35\\
1115 & 0.0259 & 15.97 & 16.4 & -0.99 & 0.62 & 1.6 & 3.69 & 1.47 & 9.62 & 4.35\\
24973 & 0.0285 & 16.16 & 16.99 & -0.06 & 2.89 & 5.59 & 16.37 & 3.16 & 9.97 & 4.35\\
1221 & 0.0345 & 17.16 & 17.83 & -0.43 & 0.97 & 1.63 & 6.72 & 2.66 & 9.73 & 4.35\\
41904 & 0.0392 & 18.42 & 19.12 & -0.36 & 4.26 & 3.49 & 29.51 & 5.18 & 8.95 & 1.0\\
\hline
\multicolumn{11}{p{11cm}}{$^\dagger$ H$_2$+He mass.}\\
\multicolumn{11}{p{11cm}}{$^\ddagger$ The conversion factor from $L_{CO\,1--0}$~[K~Km~s$^{-1}$~pc$^{2}$] to $M_{mol}$~[M$_{\odot}$], in $M_\odot$~K$^{-1}$~km$^{-1}$~s~pc$^{-2}$ units. A value of 4.35 is adopted for the normal SFG and 1.0 for galaxies classified as `mergers'. $\alpha_{\rm CO}$ includes a factor of 1.36 that accounts for the He mass in the H$_2$ phase.}\\
  \end{tabular}
\end{center}
\end{table*}

\begin{table*}[h]
\begin{center}
  \caption{\label{tab:Solomon_Howell} The combined local ULIRGs sample of \citet{Solomon97} and \citet{Howell10}.}
  \begin{tabular}{l*{9}{c}}
    ID & z & NUV & FUV & $\beta$ & S60 & S100 & $L_{IR}$           & $A_{\rm IRX}$ & $\log M_{mol}^\dagger$ \\
       &   & AB  & AB  &         & Jy  &  Jy  & 10$^{10}$ L$_\odot$& mag           & M$_\odot$ \\ 
    \hline\hline
UGC 05101 & 0.0394 & 0.0 & 0.0 & -1.05 & 11.54 & 20.23 & 1.08 & 6.78 & 9.67\\ 
IRAS F10565+2448 & 0.0432 & 0.0 & 0.0 & 0.11 & 12.12 & 15.13 & 1.19 & 7.49 & 9.75\\ 
UGC 08387 & 0.0234 & 0.0 & 0.0 & -1.69 & 15.44 & 25.18 & 0.48 & 5.42 & 9.58\\ 
Arp 220   & 0.0182 & 0.0 & 0.0 & -0.93 & 103.8 & 112.4 & 1.64 & 8.05 & 9.85\\ 
NGC 6240  & 0.0244 & 0.0 & 0.0 & -2.0 & 22.68 & 27.78 & 0.69 & 5.28 & 9.9\\ 
IRAS F23365+3604 & 0.0646 & 0.0 & 0.0 & 1.66 & 7.09 & 8.36 & 1.53 & 6.35 & 9.93\\ 
\hline
\multicolumn{9}{p{11cm}}{$^\dagger$ H$_2$+He mass. Assumes $\alpha_{\rm CO}=1.0$}.\\
\end{tabular}
\end{center}
\end{table*}

%%%%%%%%%%%%%%%%%%%%%%%%%%%%%%%%%%%%%%%%%%%%%%%%%%%
\section{C. $L_{IR}$ as a SFR indicator for low-z galaxies}
\label{app:LIR_discussion}
%%%%%%%%%%%%%%%%%%%%%%%%%%%%%%%%%%%%%%%%%%%%%%%%%%%

The IRAS selected COLD~GASS sub-sample has a significantly lower SFR and SSFR (medians 7.7~M$_\odot$~yr$^{-1}$ and $2\times 10^{-10}$~yr$^{-1}$ respectively) than our z$\sim$1 sample, although these values are still on the high side for local galaxies.
The smaller young stellar population relative to evolved stars raises the question whether a sizable fraction the $L_{IR}$ that we measure is due to radiation from old stars, reprocessed by the dust and emitted in the FIR.

Several recent studies have attempted to address this issue and arrived at seemingly contradicting results, depending on the samples, wavelength range (FIR versus sub-millimeter) and methods.
For example, \citet{Smith12} studied the SEDs of galaxies selected from the H-ATLAS survey that observe galaxies in {\it Herschel}-SPIRE sub-millimeter bands (250--500~$\mu$m). 
They conclude that for galaxies with SSFR similar to our COLD-GASS subsample as much as 50\% of the 8--1000~$\mu$m luminosity can be due to dust heated by old stellar populations.
\citet{Calzetti10} found an excess in 70~$\mu$m emission when comparing to the SFR as derived from a combination of H$\alpha$ and 24~$\mu$m emission, which they attribute to heating by old stars.
\citet{Li10} studied the 70~$\mu$m luminosity as a SFR indicator and by comparing to the results of \citet{Calzetti10} concluded that 25--40\% of the total integrated 70~$\mu$m emission in their galaxies is not related to star formation.
However, the samples of \citet{Calzetti10} and \citet{Li10} were based on galaxies with typical SFR lower by an order of magnitude than the COLD~GASS sub~sample.
If the luminosity due to star formation is scaled up by an order of magnitude, the fraction of 70~$\mu$m integrated emission due to old stars would go down to a few percents.
\citet{Dominguez_Sanchez12} compared the SFR as derived from H$\alpha$ and from {\it Herschel}-PACS 100 and 160 $\mu$m, for a sample of low redshift galaxies with very similar median SFR and SSFR as our COLD~GASS subsample. They find an excellent agreement between the two SFR indicators, in particular for spiral galaxies like our COLD~GASS galaxies.
Since old stellar populations produce very little H$\alpha$, their result indicate that in their sample, very little of the $L_{IR}$ as derived from 100 and 160~$\mu$m (very similar to the method used in this paper) is due to old stellar populations.
\citet{Boquien10} studied in detail the 100 and 160~$\mu$m emission in M33 and conclude that the 100~$\mu$m emission is a good tracer of SFR, while 160~$\mu$m emission may be affected by other processes. The origin of the FIR emission from region of low SFR density is unclear and the interpretation is strongly affected by uncertainties and modeling.

Correcting the derived $L_{IR}$ to remove emission which is unrelated to star formation is therefore not trivial.
Studies based on FIR bands (60--100 $\mu$m) tend to find smaller contamination from old stellar population than studies based on sub-millimeter (250--500 $\mu$m) bands.
As \citet{Smith12} point out, sub-millimeter selected galaxies tend to be colder and dustier than FIR selected galaxies, and their $L_{IR}$ is more likely to be affected by radiation from old stars.
Diffuse dust heated by the old stellar population is expected to be much colder than dust which is heated by the young stars and will dominate the emission only at long wavelengths, in the sub-millimeter bands.
$L_{IR}$ as derived from SED fitting to the shorter FIR wavelengths, that are dominated by star formation related emission, does not include a component of cold diffuse dust.
The fitted SED may under-predict the sub-millimeter luminosity, but still provide the correct $L_{IR}$ to be converted to SFR using the bolometric consideration of \citet{Kennicutt98}.
Since for the COLD~GASS subsample we fit \citet{CE01} SEDs to the IRAS 60 and 100 $\mu$m bands and since the SFR and SSFR of these galaxies is relatively high for local galaxies, we do not apply any correction to the derived $L_{IR}$ before converting it to SFR.

%%%%%%%%%%%%%%%%%%%%%%%%%%%%%%%%%%%%%%%%%%%%%%%%%%%
\section{D. Derivation of an Analytical \IRXbeta\ Relation}
\label{app:analytical_IRX_beta}
%%%%%%%%%%%%%%%%%%%%%%%%%%%%%%%%%%%%%%%%%%%%%%%%%%%
Let us define $A_{\lambda}$ the effective attenuation at wavelength $\lambda$ as:
\begin{equation}
 L_\lambda = L_{\lambda,\, source} \times 10^{-0.4A_\lambda}
\label{eq:L_lambda definition}
\end{equation} 
$A_\lambda$ therefore represents the magnitude by which the observed luminosity $L_{\lambda}$ is attenuated with respect to the {\it integrated} luminosity at the source $L_{\lambda, \, source}$. 
The luminosities are integrated over the entire galaxy and $A_\lambda$ includes all effects of absorption and scattering.
Generally, $A_\lambda$ will be wavelength dependent and $A_{\lambda_1} \neq A_{\lambda_2}$.
As a result, observed flux ratios (colors) will differ from those ratios at the source.
Combining e.q.~\ref{eq:L_lambda definition} and \ref{eq:beta_definition}, we can write:
\begin{equation}
 \beta - \beta_0 = -0.4 \frac{ A_{\lambda_1} - A_{\lambda_2} }{ \log(\lambda_1/\lambda_2) }
\label{eq:reddening}
\end{equation}
where $\beta$ and $\beta_0$ are the observed and unattenuated UV spectral slopes, respectively.

When integrating the luminosity over an entire galaxy, we include many individual point-like sources and each may have a different attenuation value.
Let us denote the normalized distribution of attenuations of the individual sources in the galaxy as $f_{\lambda_1}(A)$, i.e. the fraction of sources suffering attenuation between $A$ and $A+{\mathrm d}A$ at wavelength $\lambda_1$.
We then apply e.q.~\ref{eq:L_lambda definition} on each source in the galaxy according to the $f_{\lambda_1}(A)$ distribution and integrate to derive the total observed (attenuated) luminosity.
The effective attenuation that converts between the total emitted (unattenuated) luminosity and the total observed luminosity is:
\begin{equation}
\begin{array}{l}
 A_{\lambda_1} = -2.5 \log \int\limits_0^{\infty}{f_{\lambda_1}(A^\prime)\times10^{-0.4A^\prime} \mathrm{d}A^\prime} \\
 A_{\lambda_2} = -2.5 \log \int\limits_0^{\infty}{f_{\lambda_1}(A^\prime)\times10^{-0.4A^\prime(\lambda_1/\lambda_2)^{-\gamma}} \mathrm{d}A^\prime}
\end{array}
\label{eq:A1,A2integral}
\end{equation}
Here we obtain $A_{\lambda_2}$ through the attenuation distribution in $\lambda_1$ with the use of the extinction law (represented by $\gamma$) to convert from $A^\prime_{\lambda_2}$ to $A^\prime_{\lambda_1}$ under the integral.
Extinction laws in the UV-to-optical wavelengths can be approximated as a power law with power $\gamma$, excluding a `UV bump' around 2170 \AA\ in some examples.
For a point source and any two wavelengths, $\gamma$ can always be defined through the relation: $A_{\lambda_1}/A_{\lambda_2} = (\lambda_1/\lambda_2)^\gamma$, even if the extinction curve is not a true power-law.

Let us now consider a general representation for $f_{\lambda_1}$.
We will assume that most of the UV sources are scattered inside a typical range of attenuation values between $A_{min}$ and $A_{min} + {\Delta}A$.
For simplicity, we will assume that the distribution is flat inside this range.
\begin{equation}
  f_{\lambda_1}(A) = \left\{ \begin{array}{l l}
                    1/\Delta A &\quad A_{min}<A<A_{min}+\Delta A \\
		    0            &\quad else \\
                   \end{array} \right.
\label{eq:slabmix_dist_app}
\end{equation}
Due to the integration in e.q.~\ref{eq:A1,A2integral}, the effective attenuation is not very sensitive to the fine details of $f_{\lambda_1}$, with a possible exception around $A \to 0$. Most distributions that are not strictly multi-modal can be approximated as $f_{\lambda_1}$ above.

We can now insert $f_{\lambda_1}$ into e.q.~\ref{eq:A1,A2integral} and obtain $A_{\lambda_1}$ the solution for effective attenuation at wavelength $\lambda_1$:
\begin{equation}
 A_{\lambda_1} = A_{min} -2.5 \log \left( \frac{ \log(e) \left( 1-10^{-0.4{\Delta}A} \right) }{ 0.4 {\Delta}A } \right)
\label{eq:A_for_slabmix_app}
\end{equation}
The solution for $A_{\lambda_2}$ can be obtained simply by multiplying $A_{min}$ and $\Delta A$ in the right hand side of the above equation by $(\lambda_1/\lambda_2)^{-\gamma}$.
The observed spectral slope $\beta$ is then obtained by inserting the above $A_{\lambda_1}$ and $A_{\lambda_2}$ into e.q.~\ref{eq:reddening}:
\begin{equation}
 \beta - \beta_0 = \frac{1}{\log(R_{1,2})} \left[ -0.4 \left( 1-R_{1,2}^{-\gamma} \right) A_{min} +  \log\left( \frac{1-10^{-0.4{\Delta}A}}{1-10^{-0.4{\Delta}A R_{1,2}^{-\gamma}}} \right) \right]
\label{eq:beta_for_slabmix_app}
\end{equation}
where $R_{1,2} = \lambda_1/\lambda_2$.
By our definitions and when using the same rest UV wavelength in $\lambda_1$ and for deriving $A_{\rm IRX}$, the two attenuations $A_{\lambda_1}$ and $A_{\rm IRX}$ are equivalent.
With equations~\ref{eq:A_for_slabmix_app} and \ref{eq:beta_for_slabmix_app} above we can plot the expected $A_{\rm IRX}$--$\beta$ relation for various $f_{\lambda_1}$ distributions, as we did in Figure~\ref{fg:beta_for_even_mix}.

%%%%%%%%%%%%%%%%%%%%%%%%%%%%%%%%%%%%%%%%%%%%%%%%%%%%%%%%
\section{E. \IRXbeta\ Relations for Archetypal Cases}
\label{app:relation_for_archtypal_cases}
%%%%%%%%%%%%%%%%%%%%%%%%%%%%%%%%%%%%%%%%%%%%%%%%%%%%%%%%

\subsection{The Uniform Obscuring Screen}
\label{app:uniform_slab}
The simplest case to consider and the one which is most commonly assumed, is a uniform obscuring screen (Figure~\ref{fg:obscurations sketch}, case {\it a}).
In this case, all sources are subjected to the same attenuation magnitude and the distribution has the form of a Dirac delta function $f_{\lambda_1}(A) = \delta(A_{\lambda_1})$.
We set $A_{min}=A_{\lambda_1}$, $\Delta A \to 0$ and the log term in e.q.~\ref{eq:A_for_slabmix} and e.q.~\ref{eq:beta_for_slabmix} goes to zero.
The $A_{IRX}$--$\beta$ relation for wavelength $\lambda_1$ becomes:
\begin{equation}
 A_{\rm screen} =  - \frac{ 2.5\log(\lambda_1/\lambda_2)}{ 1-(\lambda_1/\lambda_2)^{-\gamma} } (\beta-\beta_0)
\label{eq:A_IRX_slab}
\end{equation}
Notice that the selection of the two wavelengths that are used in the definitions does have a small impact on this relation.
The $A_{\rm IRX}$--$\beta$ relation for this case is a straight line.
The slope is determined solely by $\gamma$, and $\beta_0$ determines the intercept with the $\beta$ axis.
We stress that the association of $\gamma$ with the linear slope and $\beta_0$ with the intercept is specific to the obscuring screen geometry and may not be so in other cases.

\subsection{The Even-Mix}
\label{app:even_mix}
Another case which is often considered is one in which the stars and dust are occupying the same slab volume with a constant densities ratio. It is represented by sketch {\it b} in Figure~\ref{fg:obscurations sketch}. In the special case of uniform densities, the attenuation distribution is:
\begin{equation}
 f_{\lambda_1}(A) = \left\{ \begin{array}{l l}
                    1/{\Delta}A &\quad 0<A<{\Delta}A \\
		    0            &\quad else \\
                   \end{array} \right.
\label{eq:even_mix_dist}
\end{equation}
When integrating the light over the entire galaxy, such a broad distribution makes some physical sense.
Stars are forming inside molecular clouds at various depths inside the clouds, not only at the core.
Older stars and stars that form near the edge of the cloud can clear away much of the obscuring material in their immediate vicinity and lower the attenuation they suffer and as a result, there is some width to the attenuation distribution.
It is also relevant for galaxies that are observed nearly edge-on if the star forming regions are scattered in the disk and the inter stellar medium contributes a significant amount of attenuation.
The expected result in these cases is a distribution of optical depths, starting from near zero and up to the maximum depths of the clouds, approximated as a step function in e.q.~\ref{eq:even_mix_dist}.

In this case we use $A_{min}=0$ in e.q.~\ref{eq:A_for_slabmix}:
\begin{equation}
 A_{\lambda_1} = -2.5 \log \left( \frac{ \log(e) \left( 1-10^{-0.4{\Delta}A} \right) }{ 0.4 \Delta A } \right)
\label{eq:A_for_even_mix}
\end{equation}
A similar solution can be obtained for $A_{\lambda_2}$ by replacing the $\Delta A$ terms with $\Delta A(\lambda_1/\lambda_2)^{-\gamma}$ in e.q.~\ref{eq:A_for_even_mix}.
For $\Delta A >> 1$ the effective attenuation increase like $A_{\lambda_1} \sim \log(\Delta A)$.
The even-mix distribution can only produce very blue (i.e. low) $\beta$.
The $\beta$ value for such a distribution is given in e.q.~\ref{eq:beta_for_slabmix}. For the limit $A_{min}=0$ and $\Delta A \to \infty$, we obtain:
\begin{equation}
 \lim_{\Delta A \to \infty} \beta-\beta_0 = -\gamma
\end{equation}

\subsection{The Bimodal Distribution}
\label{app:bimodal_distribution}

Another simple case is a bimodal distribution $f_{\lambda_1} = a\delta(A_1) + (1-a)\delta(A_2)$. This is relevant for a case in which some of the stars are forming in large dense clouds, for example at the galactic center, and others in normal HII regions of low obscuration, scattered in the disk.

A bimodal distribution has three free parameters, but since we have only two flux measurements
there are an infinite number of solutions.
For simplicity, we will choose the most trivial solution that has only two free parameters:
\begin{equation}
 f_{\lambda_1} = a\delta(A_{\rm screen}) + (1-a)\delta(A\to\infty)
\end{equation}
where $a$ is the fraction of the population suffering the lower attenuation.
This distribution has one component at an `infinite' optical depth that does not contribute to the observed UV fluxes and the other at the correct optical depth to produce the observed $\beta$.
Here, `infinite' means high enough so that the observed light is dominated by the component at lower depth, even if the obscured one is much more luminous at the source. 

While such a distribution cannot be represented in terms of $A_{min}$ and $\Delta A$, it is easy to obtain the solutions for $A_{\lambda_1}$ and $A_{\lambda_2}$ from e.q.~\ref{eq:A1,A2integral}.
We get:
\begin{equation}
 A_{\lambda_1} = A_{1,{\rm screen}}(\beta) -2.5\log(a)
\end{equation}
where $A_{\rm screen}(\beta)$ is the solution for the uniform screen geometry (e.q.~\ref{eq:A_IRX_slab}).
Since $a<1$, we get $A_{\lambda_1} > A_{\rm screen}(\beta)$ and if we attempt to derive the attenuation while assuming a uniform slab geometry we will underestimate the attenuation.
On an $A_{\rm IRX}$--$\beta$ plot, an increasing obscured fraction (decreasing $a$) will appear as a vertical shift in the attenuation for a given $\beta$ value.
This distribution can produce any $\beta$ that the obscuring screen can produce and any increase in $A$ above the $A_{\rm IRX}$--$\beta$ relation for the obscuring screen.
Even very blue galaxies with $\beta \approx \beta_0$ can have significant attenuation and such a distribution can explain the galaxies with blue UV colors and relatively high attenuation ($\beta \lesssim -1.5$, $A_{\rm IRX} \gtrsim 3$).

%%%%%%%%%%%%%%%%%%%%%%%%%%%%%%%%%%%%%%%%%%%%%%%%%%%%%%%%%%%%%%%%%%%%%%%%%%%%%%%%%
%%%%%%%%%%%%%%%%%%%%%%%%%%%%%%%%%%%%%%%%%%%%%%%%%%%%%%%%%%%%%%%%%%%%%%%%%%%%%%%%%
%%%%%%%%%%%%%%%%%%%%%%%%%%%%%%%%%%%%%%%%%%%%%%%%%%%%%%%%%%%%%%%%%%%%%%%%%%%%%%%%%

\end{document}